\documentclass[%
aip,
rsi,%
amsmath,amssymb,
preprint,%
]{revtex4-1}

\usepackage{graphicx}
\usepackage{dcolumn}
\usepackage{bm}
\usepackage{booktabs}
\usepackage{multirow}
\usepackage{xcolor}
\usepackage[inline, shortlabels]{enumitem}
\usepackage{physics}
\usepackage{amsmath}
\usepackage{xr}
\usepackage{cleveref}
\usepackage{tabularx}
\usepackage{longtable}
\usepackage[caption=false]{subfig}

\newcommand{\altetrahedral}{$\text{Al}(\text{OH})_{4}^{-}$ }
\newcommand{\superaltetrahedral}{$\text{Al}(\text{OH})_{4}^{-}\cdot14\text{H}_2\text{O}$ }
\newcommand{\aloctahedral}{$\text{Al}(\text{H}_{2}\text{O})_{6}^{3+}$ }
\newcommand{\suprealoctahedral}{$\text{Al}(\text{H}_{2}\text{O})_{6}^{3+}\cdot12\text{H}_2\text{O}$ }
\newcommand{\water}{$\text{H}_{2}\text{O}$ }
\newcommand{\al}{$\text{Al}^{3+}$ }  
\newcommand{\alisot}{$^{27}\text{Al }$}

\begin{document}

\title{\alisot NMR chemical shift of \altetrahedral calculated from first principles: Assessment of error cancellation in chemically distinct reference and target systems}

\author{Ernesto Martinez-Baez}
\email[Corresponding author: ]{e.martinezbaez@wsu.edu}
\affiliation{Department of Chemistry, Washington State University, Pullman, Washington 99164, United States}
\author{Rulin Feng}
\affiliation{Department of Chemistry, Washington State University, Pullman, Washington 99164, United States}
\author{Carolyn I. Pearce}
\author{Gregory K. Schenter}
\affiliation{Pacific Northwest National Laboratory, Richland, Washington 99352, United States}
\author{Aurora E. Clark}
\email[Corresponding author: ]{auclark@wsu.edu}
\affiliation{Department of Chemistry, Washington State University, Pullman, Washington 99164, United States}
\affiliation{Pacific Northwest National Laboratory, Richland, Washington 99352, United States}
\email{auclark@wsu.edu}

\begin{abstract}
Predicting accurate NMR chemical shieldings relies upon cancellation of different types of error in the \textit{ab initio} methodology used to calculate the shielding tensor of the analyte of interest and the reference. Often the intrinsic error in computed shieldings due to basis sets, approximations in the Hamiltonian, description of the wave function, and dynamic effects, is nearly identical between the analyte and reference, yet if the electronic structure or sensitivity to local environment differs dramatically, this cannot be taken for granted. Detailed prior work has examined the octahedral trivalent cation \aloctahedral, accounting for \textit{ab initio} intrinsic errors. However, the use of this species as a reference for the chemically distinct tetrahedral anion \altetrahedral requires an understanding of how these errors cancel, in order to define the limits of accurately predicting \alisot chemical shielding in \altetrahedral. In this work, we estimate the absolute shielding of the \alisot nucleus in \altetrahedral at the coupled cluster level (515.1 $\pm$ 5.3 ppm). Shielding sensitivity to the choice of method approximation and atomic basis sets used has been evaluated. Solvent and thermal effects are assessed through ensemble averaging techniques using ab-initio molecular dynamics. The contribution of each type of intrinsic error is assessed for the \aloctahedral and \altetrahedral ions, revealing significant differences that fundamentally hamper the ability to accurately calculate the \alisot chemical shift of \altetrahedral from first principles.
\end{abstract}


\maketitle


\section{Introduction}
\label{sec:intro}
First-principles calculations of the nuclear magnetic properties of metals in complex solutions have proven fundamental to elucidate speciation and molecular environment. Computational protocols typically employ isolated molecules of the relevant equilibrium structure, previously obtained either from quantum chemical optimizations or from experimental data. Except for small molecules in the gas phase (where absolute shieldings are accessible experimentally), the chemical shift ($\delta$) is the \textit{de-facto} parameter used for comparisons between experimental and theoretical nuclear shielding properties. Determining chemical shifts theoretically follows from the computation of the shielding ($\sigma$) of the desired nucleus ($K$) in two molecular environments (i.e reference ($\sigma_{K,ref}$) and target ($\sigma_{K,i}$) molecular configuration) using the following relationship:
\begin{equation}
    \delta_{K,i} = \frac{\sigma_{K,ref} - \sigma_{K,i}}{1 - \sigma_{K,ref}}
    \label{eqn:chem_shift}
\end{equation}\par
Agreement between experimental and theoretically derived chemical shifts relies heavily upon cancellation of errors between calculated $\sigma_{K,ref}$ and $\sigma_{K,i}$ for a given theoretical approximation. 
Previous work has demonstrated the importance of similar charge and/or geometrical environments for error cancellation in chemical shift computations. For example, the theoretically calculated chemical shifts of some Pt anionic complexes show system-dependent solvent signatures that makes error compensations ineffective.\cite{truflandier2011solvent, truflandier2010probing}       
\par
Calculated shielding constants are highly sensitive to virtually all of the approximations made in the computational protocol (i.e level of theory, basis set truncation, the approximation of electron correlation and relativistic effects). Solvent-induced medium-range effects can also have a significant impact on the shielding response of the NMR-targeted nucleus; thus should not be ignored via gas phase simulation, and may not be compensated for by simplified continuum approximations. 
As such, predictive accuracy of magnetic properties requires taking into account all the aforementioned effects, including the dynamics of the local and perhaps semi-local molecular environment (by vibrationally averaged or \textit{ab initio} dynamic approaches).
\alisot NMR spectroscopy of \al species in solution is a good example of a system where the standard NMR reference compound, \aloctahedral, presents a significantly different molecular environment from the Al-targeted species (i.e  \altetrahedral, $\text{Al}_2\text{O(H}_2\text{O})_6^{2-}$ or $\text{AlOOH}^-$).\cite{sipos2009structure}

Understanding the molecular speciation of Al as it transitions from the solution to solid phase is essential for the development of improved industrial metal processing strategies to mitigate environmental impact. In previous experimental studies, \alisot NMR spectroscopy has been used to provide direct information on  changes in Al speciation in solution, in order to understand critical aspects of Al processing, such as unexpectedly slow precipitation kinetics.\cite{graham2018situ} Therefore, this is an important system and serves as a worst case scenario for the analysis of theoretical NMR shielding sensitivity to the hierarchical model and method approximations when the reference and target molecular species are chemically dissimilar.       
There have been many previous theoretical studies on the structure and spectroscopic properties of \al species in both acidic and alkaline media.\cite{lubin2000ab, sillanpaa2001computational, sipos2009structure, pouvreau2018ab, graham2018situ} In acidic media, \al is solvated by six water molecules forming a 3+ charged cation with surrounding waters in an octahedral arrangement (Oh). In contrast, the dominant species present in caustic Al solutions is widely accepted to be the tetrahedrally (Td) coordinated aluminate ion, \altetrahedral. The calculation of $^{27}\text{Al}$ NMR shielding in the aluminate anion has been studied previously using HF, MP2 and DFT levels of theory.\cite{sykes1997molecular, kubicki1999ab, tossell1998effects,qian2009assessment} In this work, the performance of a given QM method to yield reliable NMR data was assessed by comparing calculated chemical shifts of $^{27}\text{Al}$ in \altetrahedral models (using Equation \ref{eqn:chem_shift} with $\sigma_{Al,ref}$ from the \aloctahedral model) with experimental measurements.\cite{qian2009assessment, wang2012errors}\par 
\begin{figure}[t]
\includegraphics[width=0.24\linewidth]{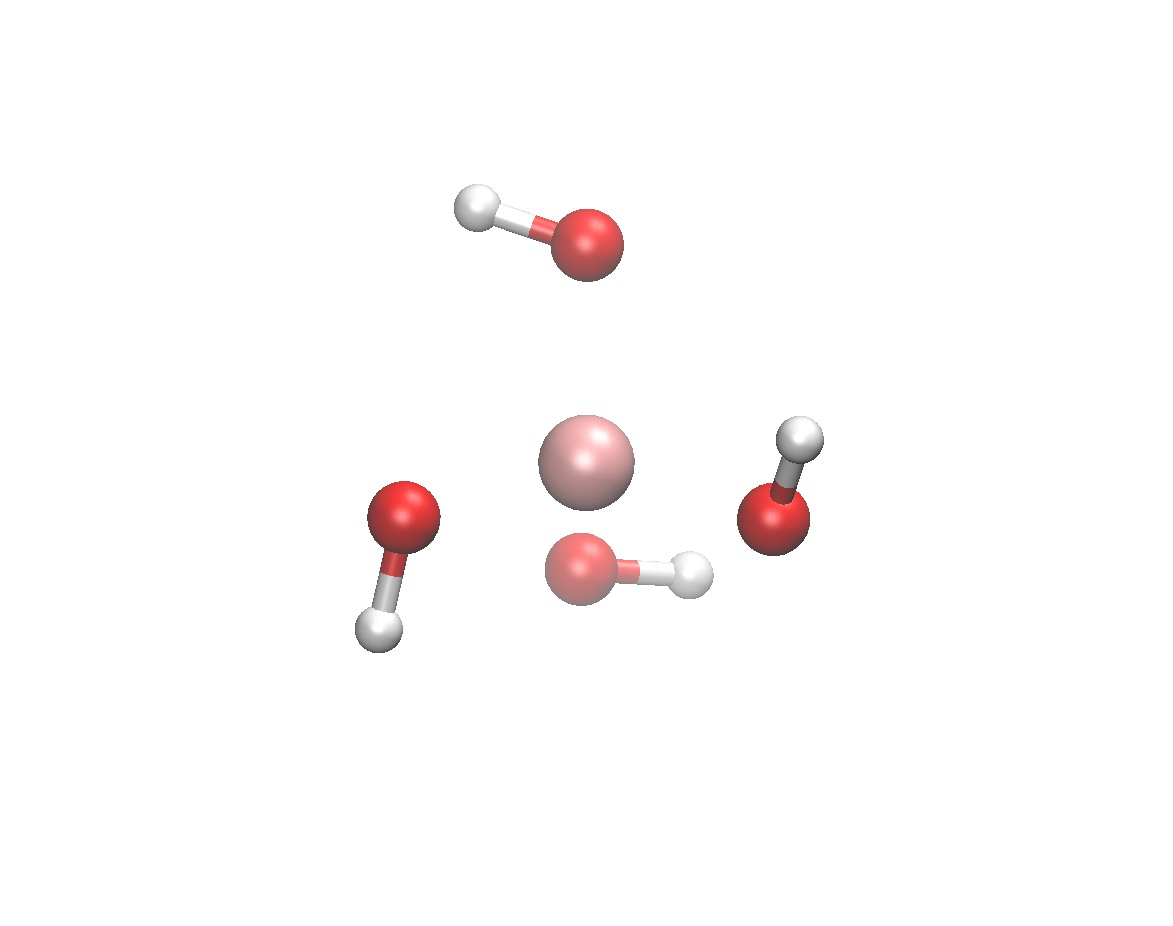}
\hspace{.005\linewidth}
\includegraphics[width=0.24\linewidth]{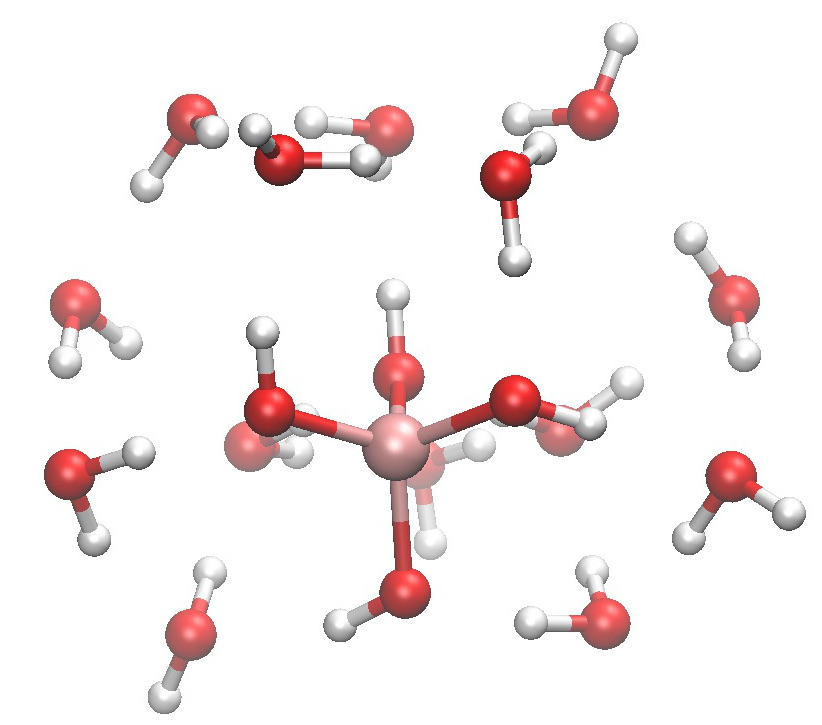}
\hspace{.4\linewidth}
\includegraphics[width=0.24\linewidth]{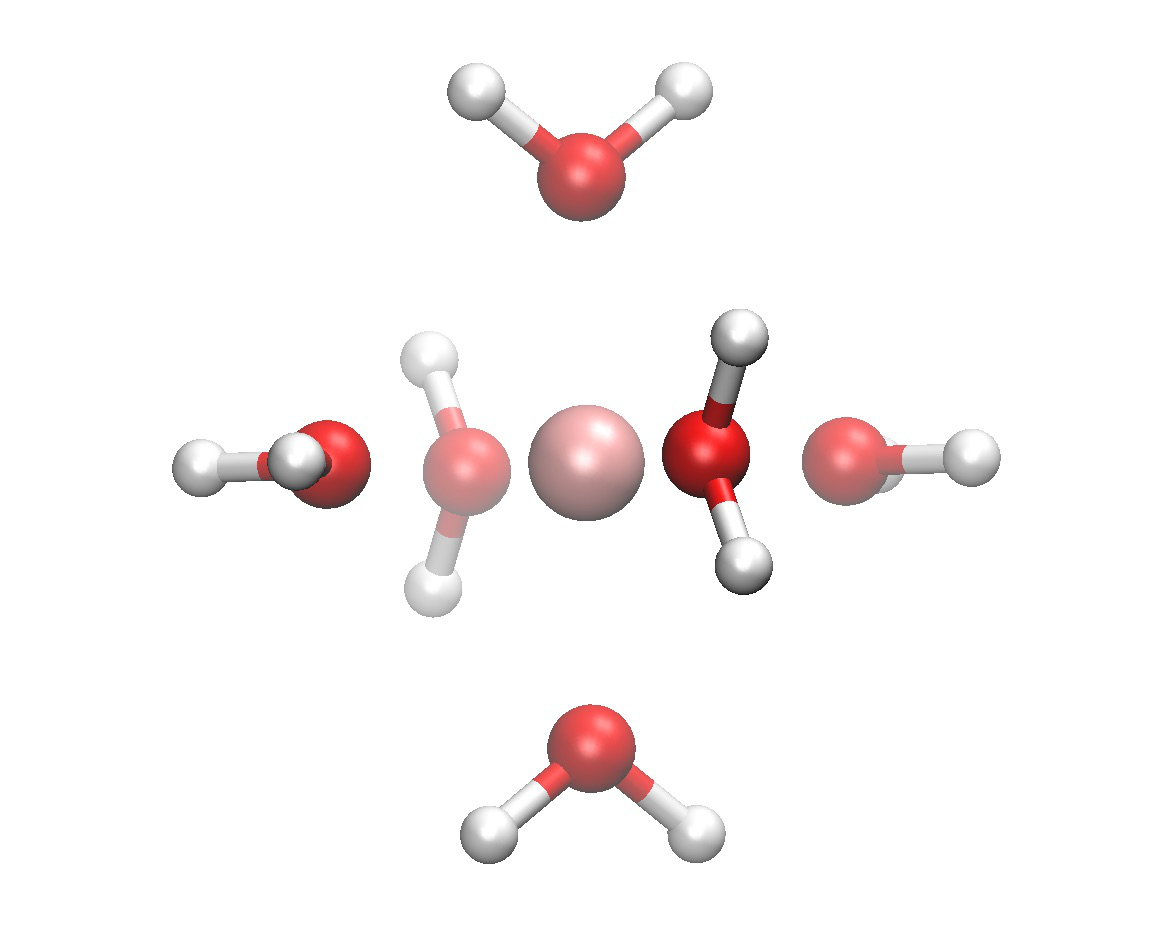}
\hspace{.005\linewidth}
\includegraphics[width=0.24\linewidth]{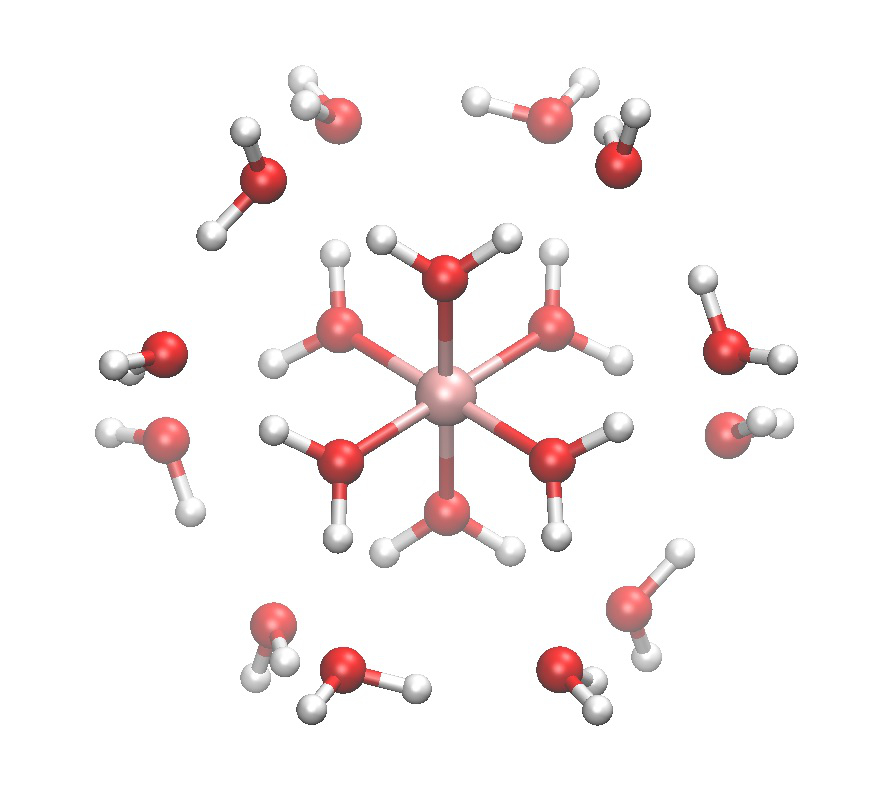}
\caption{\label{fig:geom_models} Optimized structures of gas and supermolecule model clusters at CCSD and B3LYP levels respectively with cc-pVTZ basis set. Top and bottom figures represent \altetrahedral and \aloctahedral geometries.}  
\end{figure} 
Despite these studies, a critical assessment of NMR spectroscopy data from the perspective of absolute shielding computations is still necessary to advance our understanding of error cancellations. This assessment will allow theoretical chemical shifts to be confidently compared with experimental data, and will demonstrate the potential for chemical models to predict, and account for, the significant interactions that influence experimental NMR observables. 
Recently, the Al NMR shielding constant of the \aloctahedral solvated cluster was evaluated with high-level electron-correlated coupled cluster methods using the GIAO method and Dunning-type basis sets in a layered fashion employing core-valence correlation consistent basis sets (cc-pCV$n$Z, $n=$ T, Q, 5) for the metal ion and cc-pV$n$Z set for the surrounding water molecules ($n=$ D, T, Q).\cite{antuvsek2015absolute} Corrections due to relativistic, solvent and dynamical effects were included. The most important aspect of the computational protocol, which has been frequently neglected, was the determination of second solvation shell and dynamic effects. Even though the $^{27}$\aloctahedral shielding scale was established,\cite{antuvsek2015absolute} one might anticipate an increase in the importance of electron correlation effects for an 
Al anionic analyte relative to the cationic \aloctahedral that is not cancelled out in chemical shift computations. Further, the role of solvation in the accurate determination of shielding constants may vary significantly due to the hydrogen bonding interactions of solvating waters to the terminal groups (i.e -OH groups of \altetrahedral). \par
In this work, we present a highly-accurate calculation of the Al shielding in \altetrahedral at the dilute limit. Effects of geometry, basis set, and wave function model in the NMR shielding of \al in the aluminate species, are evaluated. Corrections to gas phase non-relativistic NMR shielding constants are investigated following the additive approximation presented in equation \ref{eqn:total_shielding}. Structural dynamic corrections are calculated from ab-initio molecular dynamics (AIMD) data, as well as solvation effects from both cluster and ensemble approaches. We compare and contrast the magnitude of the aforementioned effects on the \alisot shielding in \altetrahedral ($\sigma_{Td}$) and in the standard Al NMR reference system, the hexaqua-complex species ($\sigma_{Oh}$), in order to evaluate its impact on error cancellation in Al chemical shift computations. These data demonstrate significant variations in the intrinsic errors that contribute to the two chemical systems, and highlight the fundamental limitations to the chemical accuracy of predicted \alisot shielding constants of \altetrahedral.

\section{Computational Methods}
\label{sec:methods}

\bgroup
\begin{table}[t!]
\caption{Structural parameters and HF/pCVQZ.TZ \alisot NMR shielding constants of CCSD and B3LYP optimized models of \aloctahedral and \altetrahedral species}
    \label{table:geom_models}
    \centering
\begin{tabularx}{\textwidth}{c ccccc c }
\hline
\hline
    & \multicolumn{4}{c}{\aloctahedral} & \aloctahedral$\cdot12$\water & \aloctahedral$\cdot90$\water  \\ 
\cline{2-5} \cline{6-6} \cline{7-7} 
&\multicolumn{2}{c}{CCSD} & \multicolumn{2}{c}{B3LYP} & B3LYP\footnote{(F) and (C) refer to the explicit inclusion (`full') or exclusion (`core') of the $2^{nd}$ solvation layer in the HF NMR shielding calculations for the optimized supermolecule Al models respectively.} & MD revPBE\footnote{Shielding constant of an optimized isolated Al cluster at the same computational level as the molecular dynamic runs}  \\
&DZ & TZ &DZ & TZ & TZ & DZ(MOLOPT) \\
\cline{2-3} \cline{4-5} \cline{6-6} \cline{7-7} 
    \multirow{1}{*}{$r_{(\text{Al}-\text{O})}$} & 1.956 & 1.924 & 1.951 & 1.935 &  1.916 & 1.951  \\
\midrule
    \multirow{1}{*}{$\sigma_{HF}$} & 617.4  & 612.1 & 616.3 & 613.6 & 608.7(F)/607.9(C)  &615.7 \\
\midrule
    & \multicolumn{4}{c}{\altetrahedral} &\altetrahedral$\cdot14$\water &\altetrahedral$\cdot90$\water  \\
\cline{2-5} \cline{6-6} \cline{7-7} 
    &  \multicolumn{2}{c}{CCSD} & \multicolumn{2}{c}{B3LYP} &  B3LYP & MD revPBE  \\
    &DZ/(aug-DZ)& TZ & DZ/(aug-DZ) & TZ/(aug-TZ) &  TZ &DZ(MOLOPT)   \\
\cline{2-3} \cline{4-5} \cline{6-6} \cline{7-7} 
    \multirow{1}{*}{$r_{(\text{Al}-\text{O})}$} & 1.802/1.807 & 1.778 &  1.799/1.803& 1.782/1.782 &  1.779 &1.796  \\
\midrule
    \multirow{1}{*}{$\sigma_{HF}$} & 524.8/527.8  & 521.3 &524.7/527.4  & 522.8/524.5 &  525.5(F)/527.0(C) &524.8   \\
\hline
\hline
\end{tabularx}
\end{table}
\egroup

The NMR absolute shielding constant of \alisot in \altetrahedral species was obtained using the following additive approximation:
\begin{equation}
    \sigma_{total}=\sigma_{NR} +\Delta_{rel}+\delta_{ss}+\delta_{therm}
    \label{eqn:total_shielding}
\end{equation}
where $\sigma_{NR}$ is the non-relativistic shielding, $\Delta_{rel}$ is the relativistic correction, $\delta_{ss}$ is the electron screening effect of a discreet second solvation layer and $\delta_{therm}$ denotes the average perturbation of the shielding due to the dynamic behavior of the water-solvated \al species in a thermal bath.\par
All NMR calculations were performed using the Gauge-Independent Atomic Orbital (GIAO, also known as London orbitals) method\cite{london1937theorie, hameka1958nuclear, ditchfield1972molecular,ditchfield1974self,wolinski1990efficient} to guarantee gauge invariance. 

\textit{Geometric Considerations.} Aluminum structural models of acidic \aloctahedral and alkaline \altetrahedral species were obtained through gas phase geometry optimizations  without symmetry constrains, employing the Gaussian 09 development version (D.01) [Figure \ref{fig:geom_models}].\cite{g09} Supermolecule models including a second water solvation layer (\superaltetrahedral and \suprealoctahedral) were also optimized.

\textit{Basis Set Effects.} Convergence of Al shielding with respect to basis sets quality was evaluated at the HF theory level. A wide range of basis sets, which included Pople, standard and core-valence Dunning and Jensen types, was studied. Effects of extra polarization and diffuse functions were revealed. Furthermore, the sensitivity of the Al NMR shielding constant with respect to changes in the basis sets of surrounding solvating O and H atoms was also tested for Dunning and Jensen basis sets types. (If the cc-pCVTZ basis was used for Al and cc-pVTZ for O and H atoms, we referred to this combination as pCVTZ.TZ. Similar nomenclature is used for other combinations).\par

\begin{figure*}[t!]
\centering
\begin{tabular}{@{}ccc@{}}
    \includegraphics[width=0.49\linewidth]{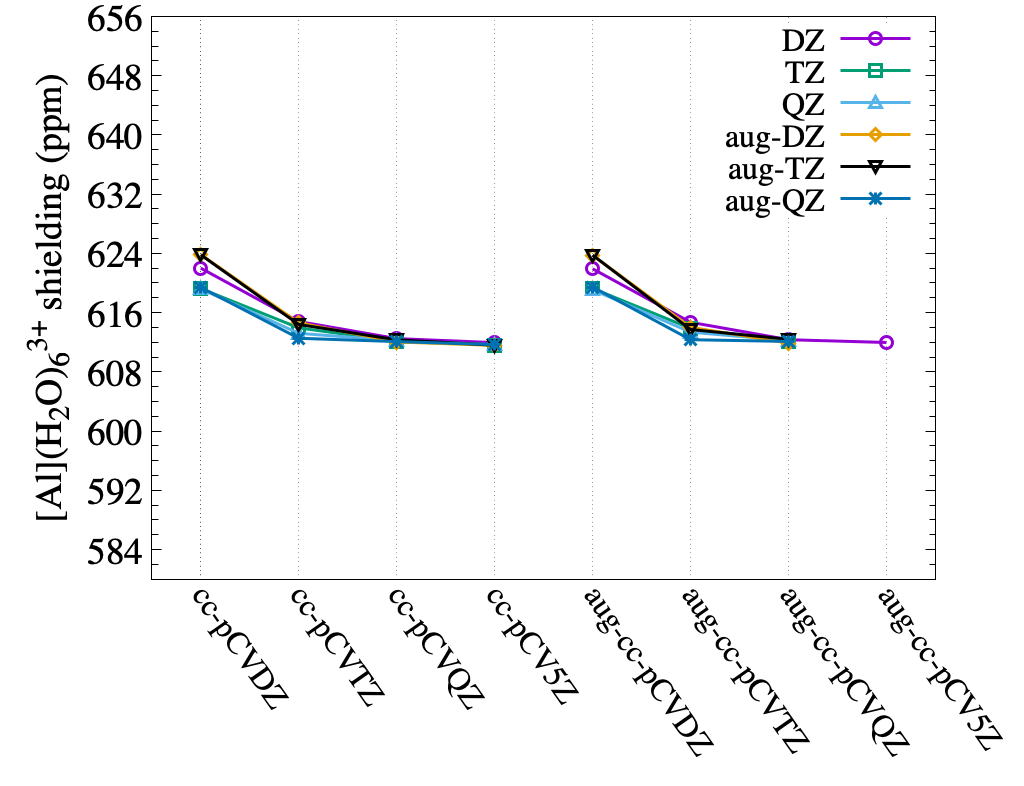} &
    \includegraphics[width=0.49\linewidth]{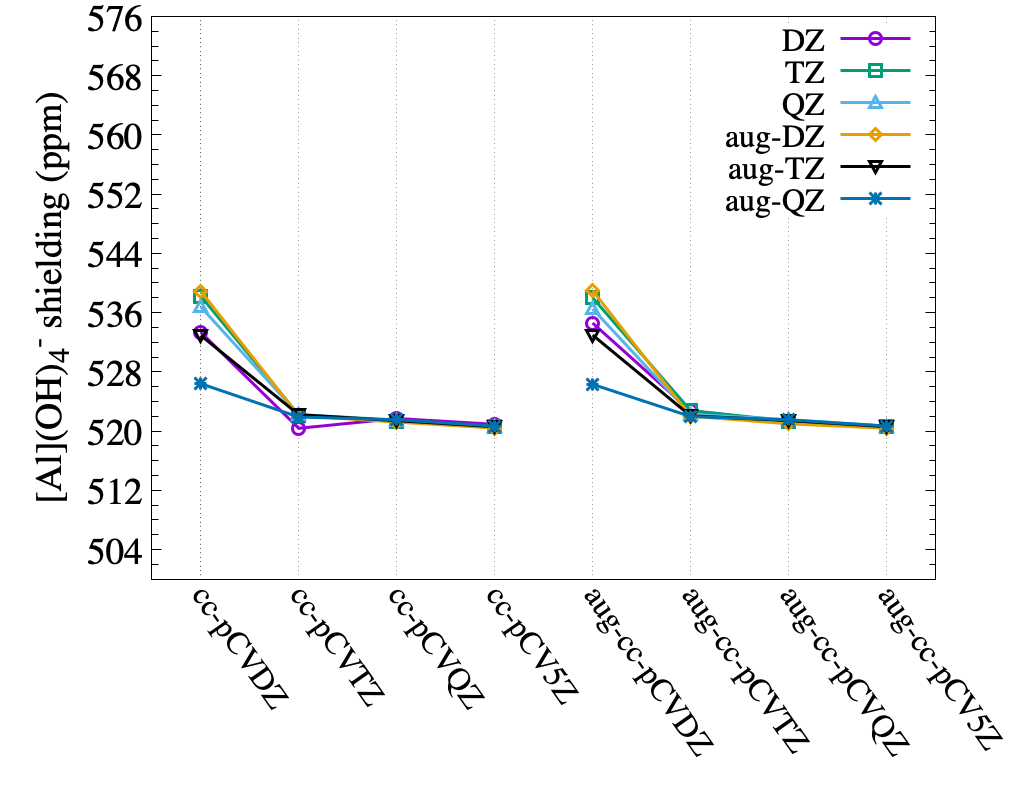} &
\end{tabular}
\caption{HF GIAO Al NMR shielding convergence behavior as a function of standard and core-valence Dunning basis sets quality. Gas phase \aloctahedral and \altetrahedral models optimized at CCSD/cc-pVTZ level}
\label{fig:bs_dunning}
\end{figure*}

\textit{Coupled Cluster Methods.} The non-relativistic shielding ($\sigma_{NR}$) of \altetrahedral was obtained at the couple cluster singles and doubles (CCSD) level using the CFOUR v2.1 software.\cite{cfour} Uncertainty in $\sigma_{NR}$ due to model geometry approximation, basis set incompleteness and truncation errors was evaluated. \par

\textit{Relativistic Effects.}  Relativistic corrections were calculated at the HF level as the difference between 4c-Dirac-Hartree-Fock (DHF)\cite{visscher1997approximate} shielding and the non-relativistic values using the Levy-Leblond\cite{levy1967nonrelativistic} Hamiltonian as implemented in DIRAC.\cite{DIRAC18}\par 

\textit{Solvent Screening Effects.} The electron screening perturbation from higher solvation layers ($\delta_{ss}$) was evaluated at the HF level as the difference between the shielding of the \al supermolecule models and the resulting first-layered solvated structures (i.e \altetrahedral or \aloctahedral inner configurations extracted from their respective supermolecule models).\par 

\textit{Thermal and Dynamic Effects.} Thermal corrections were estimated as difference of the averaged shielding from a few hundred cluster structures (\altetrahedral and \aloctahedral without extra waters) extracted from an ab-initio molecular dynamics (AIMD) trajectory, and the shielding of an optimized Al cluster with the same method employed in the AIMD runs:
\begin{equation}
\delta_{therm} =\langle\sigma[\text{Al}(\text{OH})_{4}^{-}]\rangle_{MD}-\sigma[\text{Al}(\text{OH})_{4}^{-}]_{opt}^{MD}
\end{equation}
The AIMD simulation included up to 3 layers of solvated waters around the \altetrahedral and \aloctahedral species, which enforced a more realistic dynamic framework. Consequently, $\delta_{thermal}$ represents both internal and solvation-driven dynamic effects. 
The AIMD simulations were carried out within the Born Oppenheimer framework (we indistinctly use BO-AIMD as AIMD throughout the text). All systems were initially relaxed with Classical Molecular Dynamics force fields followed by density equilibration in NPT ($\mathrm{10ps}$) before canonical ensemble (NVT) pre-equilibration ($\mathrm{5ps}$) and production runs ($\mathrm{30ps}$) using the Gaussian and Plane Wave (GPW) method within the Quickstep module of the CP2K software.\cite{lippert1997hybrid, vandevondele2005quickstep}
Periodic cubic boxes of side length 14.35 and 14.20 \r{A}  containing one \altetrahedral and \aloctahedral ion respectively surrounded by 90 discreet water molecules were used. The temperature was maintained at 300K using the CSVR\cite{bussi2007canonical} (canonical sampling through velocity rescaling) thermostat with a time constant of 20 fs. Valence electrons were treated at the DFT level employing revPBE\cite{perdew1996generalized,perdew2008restoring,zhang1998comment} functional with dual basis sets DZVP-MOLOPT-SR\cite{vandevondele2007gaussian} type with density from a cutoff of 400Ry. Core electrons on all atoms were represented by Goedecker-Teter-Hutter (GTH) pseudopotentials\cite{goedecker1996separable,hartwigsen1998relativistic,krack2005pseudopotentials}. The Grimme D3\cite{grimme2010consistent} dispersion correction for the revPBE functional was added with a 40 \r{A} cutoff. A timestep of 1 fs was used to generate the AIMD trajectories.\par
A similar computational scheme has been previously employed by Antu\v{s}ek et al. to calculate the absolute shielding of \al in \aloctahedral.\cite{antuvsek2015absolute} Their work, in combination with our results for \altetrahedral, is employed to increase the understanding of error cancellation between \al acidic and alkaline species whenever possible.


\section{Results and discussion}
\label{sec:results}
Predicted nuclear shielding values in molecular systems are highly sensitive to variations of the molecular structure around the nucleus of interest, the choice of the atomic basis sets representation, and the wave function model approximation. Herein we first examine the role of minute changes to the gas-phase geometry and then basis set at a fixed geometry at the HF level. We then report the calculation of the absolute shielding constant of aluminate including electron correlation at the CCSD level and successive corrections for thermal, solvation and relativistic effects. A point-by-point comparison of the magnitude of the shielding corrections between \aloctahedral and \altetrahedral species is performed.

\textit{The Role of Gas Phase Geometry: \altetrahedral vs \aloctahedral.} Herein the HF approximation is employed to determine the shielding constants ($\sigma_{HF}[^{27}\text{Al})]$) in \altetrahedral and \aloctahedral species, using different geometries obtained from systematically more accurate methods [Table \ref{table:geom_models}]. Structural parameters of Al geometries in both complexes are shown in Table \ref{table:geom_models} as obtained by CCSD and B3LYP optimizations with double and triple-$\zeta$ Dunning basis sets. Optimizations at the CCSD level with bigger than triple-$\zeta$ basis are restricted due to the high demands on computational resources. The supermolecule models \superaltetrahedral and \suprealoctahedral were optimized only at the B3LYP/cc-pVTZ level since the CCSD approach is inexpedient for such large number of atoms.\par

Changes in Al-O distances between CCSD and B3LYP optimized models for a given basis set framework are on the order of $10^{-3}$\AA. For example, the B3LYP/DZ $\mathrm{Al-O}$ average bond distance in \aloctahedral varies only in $\sim0.25$\% when re-optimized with CCSD/DZ. For both Al species, the triple-$\zeta$ optimizations in B3LYP gave slightly expanded $\langle \text{Al-O} \rangle$ bond lengths (+0.01 \AA) compared with the CCSD structures, which was opposite to the trend observed using a double-$\zeta$ basis. The structural changes due to method selection (CCSD vs B3LYP) are small but sufficient to vary the Al shielding constants by nearly 1.5 ppm. (i.e 1 ppm is greater than the experimental accuracy of NMR measurements). Based upon these data, \textit{the error in $\sigma(HF/pCVQZ.TZ)$ associated with geometric uncertainty in the Al cluster models, calculated as the difference between CCSD/TZ and B3LYP/TZ optimized structures, is $ \leq 2$ ppm for both \aloctahedral and \altetrahedral species}. The basis sets cardinality employed during the optimization steps had a slightly higher impact on the resultant structures and corresponding Al shieldings than the model choice (i.e CCSD or B3LYP). For example, \textit{the indirect effect of the basis sets employed during the geometry optimizations in the Al shielding constant (i.e $\sigma(\mathrm{DZ})$-$\sigma(\mathrm{TZ})$) is in the range of 2-5 ppm change} [see Table \ref{table:geom_models}].\par

\textit{Basis Set Effects at the HF Level.} In chemical shift calculations where the reference model is significantly different from the target system, understanding the convergence behavior of NMR shieldings as a function of basis sets characteristics and size proves fundamental to discern between fortuitous error cancellation and the best achievable chemical shift values computed at a chosen theory level. To the best of our knowledge, previous work on aluminate chemical shift calculations in the dilute limit has been performed almost exclusively with Pople basis.\cite{sykes1997molecular,tossell1998effects,kubicki1999ab,qian2009assessment,graham2018situ,dong201927al} Efforts to evaluate the $^{27}$\altetrahedral NMR absolute shielding and chemical shift convergence with respect to basis sets types in a rigorous way are lacking. Our extensive basis sets study has two main objectives: 
\begin{enumerate*}[(a)]
    \item to present recent reliable data of \alisot NMR HF shielding constants for both \altetrahedral and \aloctahedral as a function of basis sets type to serve as reference for future computational Al NMR studies, and 
    \item to demonstrate the comparative performance of standard basis sets families for the calculation of aluminate NMR chemical shift computations.
\end{enumerate*}\par
As expected, the magnitude and basis set convergence behavior of the \al shielding is highly sensitive to both, the close molecular environment (i.e differences between Al species) and the number and type of functions used for the atomic basis sets description. Basis set dependent HF shielding values of \alisot in the \aloctahedral species extend from 650 to 600 ppm whereas in the \altetrahedral model the shieldings range from 575 to 516 ppm ($\sim$ 70 to 100 ppm deshielding with respect to \aloctahedral).
Overall, all basis sets families except Pople and standard Dunning types showed a smooth and uniform-across-species shielding convergence behavior towards their respective complete basis set (CBS) limit with increasing basis sets size (see Figure \ref{fig:bs_dunning}, and Figure \ref{fig:bs}).
\begin{figure}[!h]
\centering
\begin{tabular}{@{}c@{}}
    \includegraphics[width=0.6\textwidth]{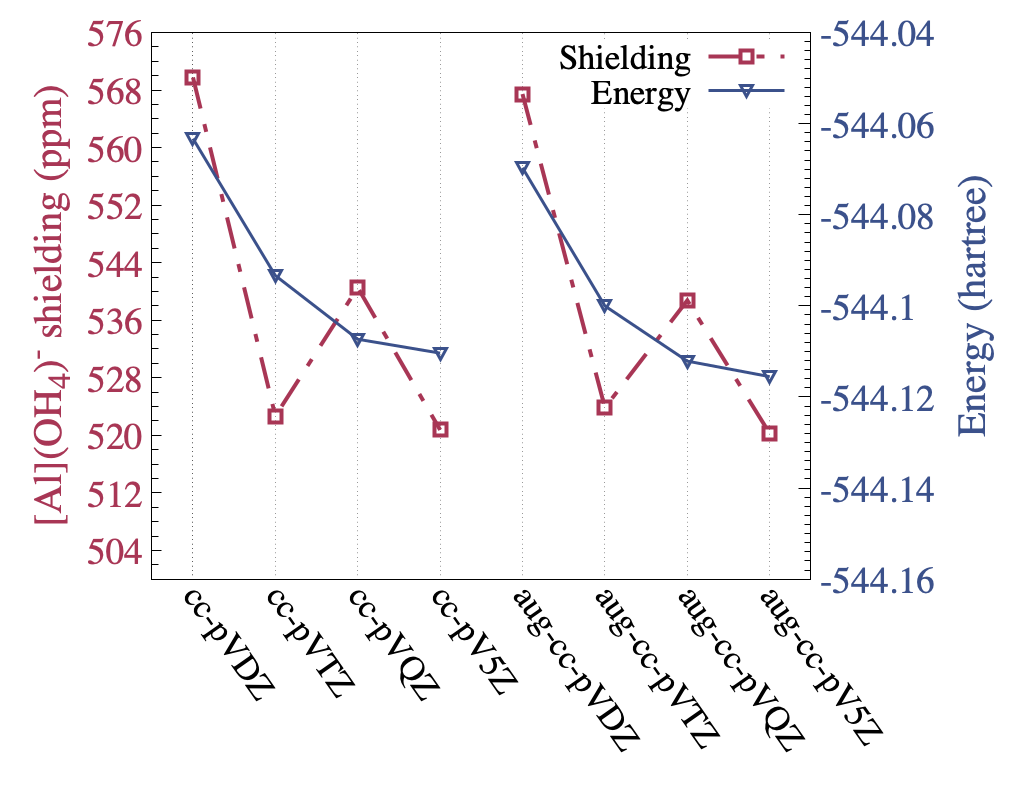} \\[\abovecaptionskip]
    \small (a)
\end{tabular}
\begin{tabular}{@{}c@{}}
    \includegraphics[width=0.6\textwidth]{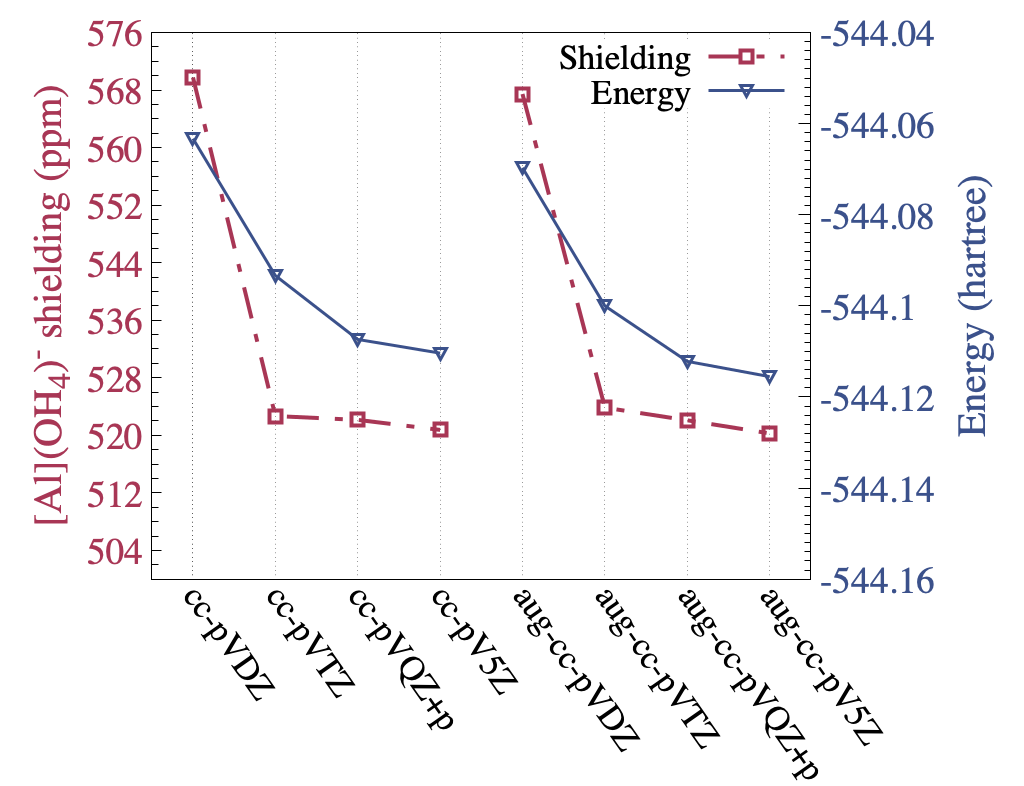} \\[\abovecaptionskip]
    \small (b)
\end{tabular}
\label{fig:bss_correction}
\caption{a) Example of $\sigma_{Al}$(ppm) and energy convergence (hartree) behavior in the \altetrahedral species for standard Dunning basis sets (cc-pVTZ for O and H atoms), and b) corrected convergence results after modification of the QZ basis sets result by addition of a p tight functions.}
\end{figure}

\textit{Use of correlation-consistent core-valence Dunning or pc-S Jensen  basis sets families for Al guarantees a monotonically decreasing, convergence behavior of $\sigma($Al$)$ towards the CBS limit as a function of increasing basis sets size for both aluminum species at a moderate computational cost}.\par
As previously noted,\cite{antuvsek2015absolute} the use of DZ quality is not recommended for shielding tensor calculations in view of their considerable deviations from the shielding CBS limit. In addition, $\sigma(^{27}Al)$ calculations at the DZ level are very sensitive to changes of the O and H basis sets size with a variability of 5 to 12 ppm units. Employing a quad-$\zeta$ basis set to describe the Al atom, guarantees shielding convergence for both \al species. At the QZ level, $\sigma($Al$)$ values become unaffected (less than 1 ppm fluctuations) by changes in surrounding O and H (i.e in water or $\text{OH}^-$ solvent molecules) basis sets size. The CBS shielding ($\sigma_{CBS}$) is calculated as:
\begin{equation}
    \sigma_{n}=\sigma_{CBS}+A(n+1)\exp(-9.03\sqrt{n})
    \label{eqn:Karton-Martin CBS}
\end{equation}
utilizing the Karton-Martin formula\cite{karton2006comment} with $n=4$ and $5$. The core-valence Dunning CBS limit of \alisot shielding constants approaches $\sim$611.5 and $\sim$520.4 ppm for \aloctahedral and \altetrahedral species respectively, which results in a calculated \altetrahedral chemical shift that converges to $\sim$ 91.1 ppm. A $\sim$ 10-12 ppm discrepancy with experiment in the calculated chemical shift of \altetrahedral can be considered as the best achievable resolution at the Hartree-Fock level with the employed models. 
Surprisingly, the standard Dunning basis sets behavior showed an odd discontinuity from the exponential decaying trend at the QZ level (Figure 3a). A closer look at the uncontracted sp correlation functions for the cc-pVTZ, cc-pVQZ, and cc-pV5Z sets for aluminum atom helped explain the oddity, it is found that the VTZ, VQZ, V5Z series does not seem to form a strictly systematic increase in the description of especially tight p functions. For the cc-pVQZ set particularly, the tightest uncontracted p exponent is more diffused than that of the cc-pVTZ set by a factor of $2.64$. This may change the convergence trend for a particular chemical property while still preserve a smooth convergence for energetic properties.
Re-balancing the systematic increase for sp uncontracted tight functions with respect to basis set size can rebuild the smooth convergence for the shield constant. In fact, simply uncontracting one more tighter p function ($\alpha=0.8706$ in this case) from the cc-pVQZ HF primitives has achieved a smooth convergence (see Figure 3b). The smooth convergence of the Al shielding when using the unchanged core-valence sets (cc-pCV$X$Z) also stresses the importance of uncontracted tight sp functions for properties sensitive to core electrons description, as they were designed to describe the outer-core electron correlation. 

\textit{Coupled Cluster Calculated Shielding of \altetrahedral in the Gas Phase.} The absolute \alisot shielding in \altetrahedral was evaluated at the CCSD level employing the pCVQZ.TZ basis set. We assume that the changes in basis sets dependency for bigger than pCVQZ.TZ basis sets at the CCSD level will remain within 1-2 ppm,\cite{antuvsek2015absolute} hence using pCVQZ.TZ will suffice to obtain reliable shieldings with a reasonable computational cost. Electron correlation affects the \alisot shielding constants significantly in both species. \textit{CCSD correlation contribution constitutes 4.1\% of the Hartree-Fock value for \altetrahedral, which is almost double the effect on \aloctahedral (2.5\%)}. The absolute decrease associated with the \altetrahedral shielding constant is 5.6 ppm units greater than the one for \aloctahedral (Table \ref{table:nmr_ccsd}).\cite{antuvsek2015absolute}.\par
Errors in $\sigma_{NR}^{Td}$ associated with geometry uncertainty in the model (CCSD/pCVQZ.TZ shielding differences between B3LYP and CCSD optimized models using cc-pVTZ during optimization), basis set incompleteness (shielding difference evaluated at CCSD/pCVQZ.TZ and CCSD/pCVTZ.TZ) and model truncation (CCSD(T) compared with CCSD shielding at the pCVQZ.TZ level) were each smaller than 4 ppm and of similar magnitude to the ones obtained for the Al shielding constant in the \aloctahedral species,\cite{antuvsek2015absolute} (Table \ref{table:uncert}).   
The average uncertainty of $\sigma_{NR}$ at the CCSD/pCVQZ.TZ is calculated propagating individual approximation errors using $e_{T} = \sqrt{\sum_{i}^{3}e_i^2}$. \textit{Henceforth, the absolute non-relativistic \al shielding constant in \altetrahedral is approximated to the CCSD/pCVQZ.TZ value of $499.6 \pm 5.3$ ppm}.
Taking into account electron correlation contributions exacerbates the differences between the calculated and experimental chemical shift values of \altetrahedral. The system-specific magnitude of electron correlation perturbations reinforces the need to explore medium-to-long range solvation effects for error cancellation in \alisot chemical shift calculations.\par
\begin{table}[!t]
\caption{Non-relativistic CCSD/pCVQZ.TZ shielding constants (in ppm) of Al in \altetrahedral clusters optimized at the CCSD method with triple-$\zeta$ basis set.}\label{table:nmr_ccsd}
\begin{tabularx}{0.8\textwidth}{XXX XXX XX}
\hline
\hline
            \multicolumn{3}{c}{\aloctahedral\footnote{$^{27}Al$ shielding values for \aloctahedral are taken from \textcite{antuvsek2015absolute}}} & \multicolumn{3}{c}{\altetrahedral} & \multicolumn{2}{c}{$\delta($\altetrahedral$)$} \\ 
            \cline{1-3} \cline{4-6} \cline{7-8}
HF  &  CCSD & $\Delta$ & HF  & CCSD & $\Delta$ & HF & CCSD \\ 
614.6   & 598.7 & 15.9 & 521.1 & 499.6  & 21.5 & 93.5 & 99.1\\
\hline
\hline
\end{tabularx} 
\end{table}
\begin{figure}[b!]
    \centering
    \includegraphics[width=0.6\linewidth]{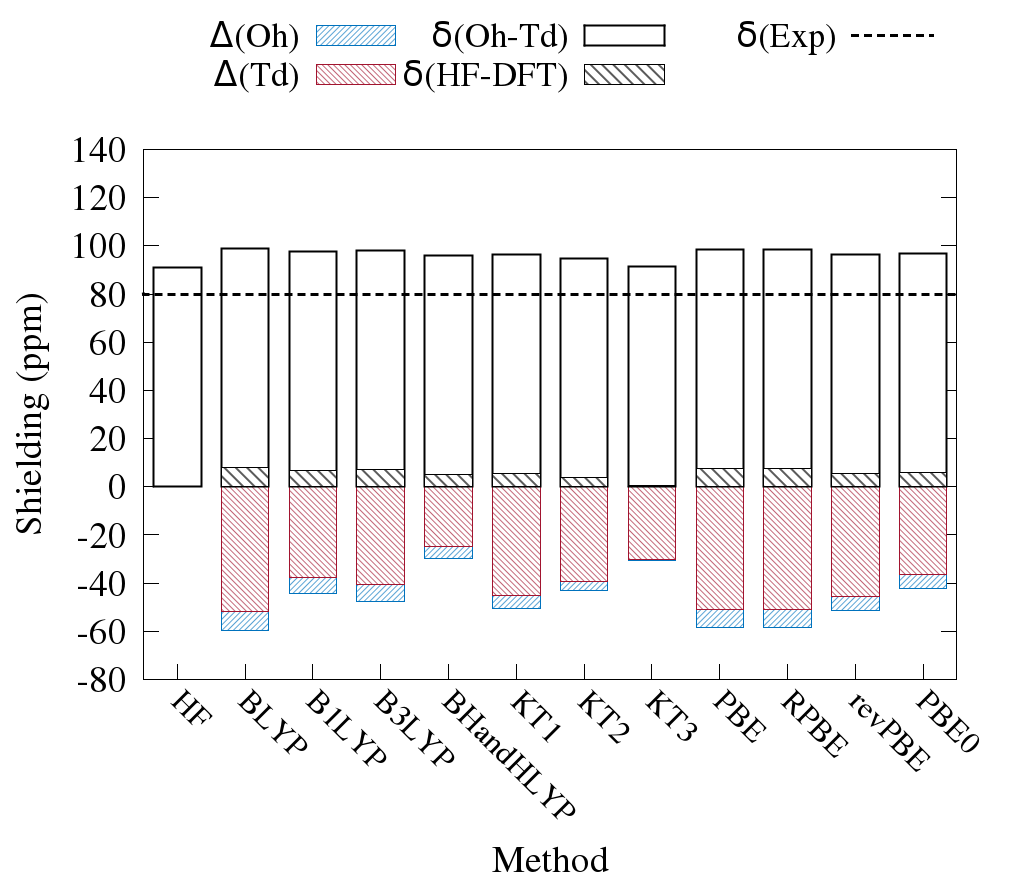}
    \caption{DFT Al NMR shielding constants (in ppm) calculated with the pCVQZ.TZ basis set combination. Geometries were optimized at the CCSD/cc-pVTZ level. $\Delta$ values are calculated as $\sigma_{DFT}-\sigma_{HF}$.}
    \label{fig:dftl}
\end{figure}
Electron correlation effects on Al NMR shieldings were also explored using DFT approximation. Concurring with the CCSD electron correlation deshielding effect, \textit{all \al DFT shielding constants are shifted towards lower values with respect to the HF magnitude for both Al species} (Figure \ref{fig:dftl}). There is a significantly smaller deshielding effect in the \al shielding of the \altetrahedral species compared with the hexa-aqua complex for almost all DFT methods studied here, with the exception of the KT3 functional. The differences in the electron correlation effects at the DFT level results in an overall increase of the \altetrahedral chemical shift. Hence, the \altetrahedral Al NMR chemical shift calculated using common DFT functionals deviates downfield in approximately 10 ppm from the 80 ppm experimental value (see Table \ref{table:nmr_dft} for numerical values). 

\textit{Relativistic Effects and the Shielding Constant.} Relativistic effects on the shielding were obtained through a Dirac-Coulomb Hartree-Fock ground state shielding tensor calculation using simple magnetic balance in conjunction with London orbitals\cite{olejniczak2012simple, aucar1999origin, pecul2004electric} and a point charge nuclear model.
We employed uncontracted relativistic atomic natural orbital basis sets ANO-RCC\cite{roos2004main} for aluminum and cc-pVTZ for O and H atoms. \textit{The relativistic correction to the \al shielding is insensitive to the chemical environment given by the similarity of $\Delta_{rel}=12.0$ ppm for \altetrahedral compared to the 12.6 ppm for \aloctahedral}.\cite{antuvsek2015absolute} \par 

\begin{figure}[!htb]
\centering
\begin{minipage}{\linewidth}
\includegraphics[width=0.6\linewidth]{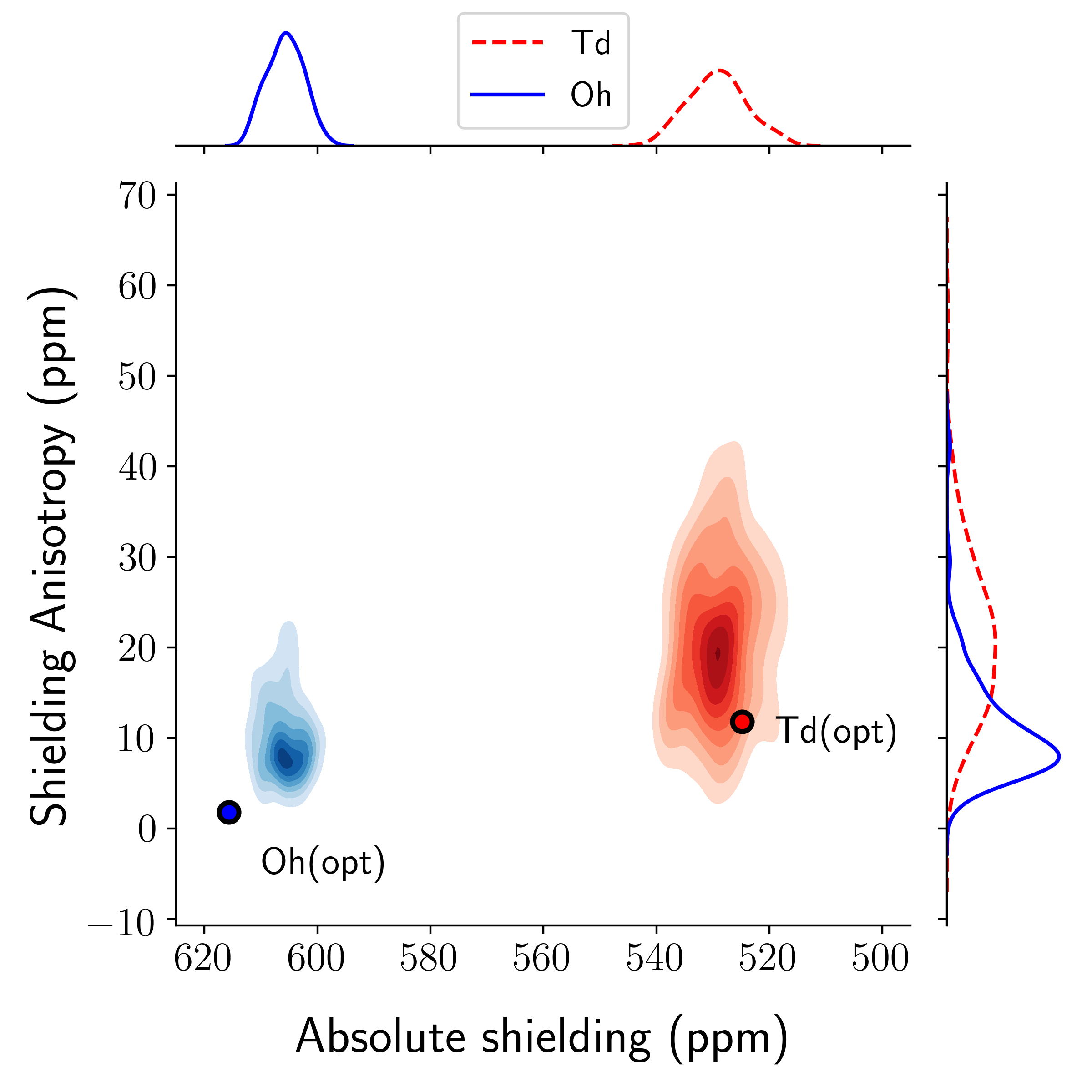}\par
\end{minipage}
\caption{Ensemble analysis of \altetrahedral (red) and \aloctahedral(blue) shielding constants from AIMD simulations. Top panel: Kernel density estimation (KDE), using a gaussian kernel, of the isotropic shielding constant distribution. Right panel: KDE of the anisotropy of the shielding tensor. Center panel shows the bivariate KDE heat maps of both Al species. Annotated colored dots on the center panel represent the shielding of \aloctahedral (blue) and \altetrahedral(red) clusters optimized in vacuum employing the same method used for the AIMD run.}
\label{fig:distribution}
\end{figure}
 
\textit{Electrostatic Screening by a Fixed Solvation Shell.} Solvent screening effects ($\delta_{ss}$) were obtained employing the supermolecule models (\superaltetrahedral and \suprealoctahedral) (see Section \ref{sec:methods}). 
Octahedral and tetrahedral \al first solvation shells of the supermolecule models show distinct overall symmetrical arrangements of their first-solvation-layer H-atoms in order to maximize hydrogen bonding interaction with second layer water molecules. Reorganization of the first solvation layer of the cationic and anionic \al species by the addition of a second solvation layer in the optimization process, leads to significant contrasting changes in the \alisot shieldings of \altetrahedral and \aloctahedral. This is seen by comparing the shieldings of the first-layered solvated Al structures (see Table \ref{table:geom_models} supermolecule shieldings labeled (C)) with gas phase models at the B3LYP/cc-pVTZ. For example, there is a $\sim 6$ ppm deshielding in the hexaqua complex and a $\sim 4$ ppm shielding in aluminate. Explicit inclusion of the $2^{nd}$ solvating waters in the shielding calculation results in a dampening of the structural deformation effect of 0.8 ppm for \aloctahedral and -1.5 ppm for \altetrahedral. The latter solvent-induced perturbations of the shielding are labeled as static solvent screening effect ($\delta_{ss}$).   
The recalculated Al chemical shift from the \aloctahedral and \altetrahedral structures extracted from the supermolecule clusters is 80.8 ppm ($\sim$1 ppm error from experimental aluminate NMR). It is clear that, \textit{employing an explicit second solvation layer to optimize the two \al species alters the internal structure of the first solvation layer in a way that maximizes error cancellation in aluminate chemical shift computations at the HF level}. Notwithstanding, the above fortuitous error cancellation that leads to excellent theory-experiment agreement is not a direct measure of accuracy of each \al species' absolute NMR shielding calculation. A more realistic picture of the averaged complex solvation environment of these ions should consider averaged dynamical solvation effects not included in $\delta_{ss}$.\par   
\textit{Dynamic Solvation Effects upon Shielding Constants.} We evaluated solvent-including thermal effects on the shielding constant of both Al species from first-principles molecular dynamics simulations at the dilute limit (see details in Section \ref{sec:methods}).  Statistical ensemble averaging of the shielding was performed from 150 and 200 structures for \aloctahedral and \altetrahedral respectively. Aluminate and \aloctahedral clusters were extracted every 50 and 20 fs respectively. The chosen sub-ensemble sample size guaranteed convergence of the averaged shielding constant for each species. $\delta_{therm}$ convolutes changes of Al-O bond lengths and O-Al-O angles due to intramolecular motions and intermolecular interactions with surrounding water molecules. The calculated HF/pCVQZ.TZ  AIMD ensemble-averaged shielding constants were 529.2 $\pm 0.7$ and 605.7 $\pm$ 0.5 ppm for the aluminate and \aloctahedral species respectively (confidence intervals were calculated from the standard error of the mean as $t\cdot\frac{\sigma_{std}}{\sqrt{N}}$, $t=1.96$). \textit{For \aloctahedral the $\delta_{therm}$ correction is negative ($-10$ ppm) whereas for \altetrahedral it becomes a shielding effect ($+5.2$ ppm)}. Figure \ref{fig:distribution} shows the ensemble of calculated shielding constants for both aluminum systems against the shielding anisotropy ($\sigma_{aniso}$): 
\begin{equation}
    \sigma_{aniso} = \sigma_{zz} - \frac{\sigma_{xx} + \sigma_{yy}}{2}
\end{equation} 
where $\sigma_{xx}, \sigma_{yy}$ and $\sigma_{zz}$ are the diagonal elements of the shielding tensor matrix respectively. 
Definite structure-to-property correlations are difficult to assess in terms of isotropic/anisotropic shielding fluctuations. Nevertheless, a wider spread of the iso-aniso shielding density map is correlated to a greater fluctuation of the electronic density around each shielding tensor component which in turn could indicate a more complex structural dynamics. The shielding anisotropy of \altetrahedral extends to 40 ppm, which is considerably wider than the one of the \aloctahedral ensemble with a maximum of 23 ppm. This is indicative of a more heterogeneous electron density fluctuation around the \al center in the solvated aluminate system compared to the hexa-aqua Al species. Analysis of direct correlations between internal molecular motions, electron density fluctuations and nuclear shielding variability will be the focus of a future study.\par   
\begin{table}[!t]
\centering
\caption{Summary of the corrections to the HF/pVCQZ.TZ shielding constant of \al in \altetrahedral species (in ppm). Aboslute shielding of \alisot in \altetrahedral at the CCSD/pCVQZ.TZ level ($\sigma_{total}$)}
\label{table:nmr_correc}
\begin{tabularx}{0.75\textwidth}{ X@{\hspace{0.5cm}} c@{\hspace{0.5cm}}c@{\hspace{0.5cm}}c@{\hspace{0.5cm}}c@{\hspace{0.5cm}}c@{\hspace{0.5cm}}c@{\hspace{0.5cm}}}
\hline 
\hline 
Ion  &  $\Delta_{corr}$ & $\Delta_{rel}$ & $\delta_{therm}$  & $\delta_{ss}$ & $\Delta_{total}$ & $\sigma_{total}$  \\ 
\cline{2-2} \cline{3-3} \cline{4-4} \cline{5-5} \cline{6-6} \cline{7-7}
\altetrahedral   & $-21.7 \pm 5.3$ & 12 & 5.2 & $-1.5$  & $-5.99$ & 515.13 $\pm$ 5.3  \\
\hline
\hline
\end{tabularx} 
\end{table}
\textit{Absolute \alisot NMR shielding in \altetrahedral. Comparison of shielding corrections between \altetrahedral and \aloctahedral.}
Using equation \ref{eqn:total_shielding} we can approximate the total shielding constant of \alisot in \altetrahedral to be 515.13 $\pm$ 5.3 ppm. With equation \ref{eqn:chem_shift} and the reported absolute scale of \al ion shielding in \aloctahedral of 600.0 $\pm$ 4.1,\cite{antuvsek2015absolute} we can estimate the chemical shift of aluminate to be 84.87 $\pm$ 6.7 (the confidence interval was obtained through error propagation), which falls within the interval of the experimental measurement.
\begin{figure}[h]
\centering
\begin{minipage}{\linewidth}
\includegraphics[width=0.6\linewidth]{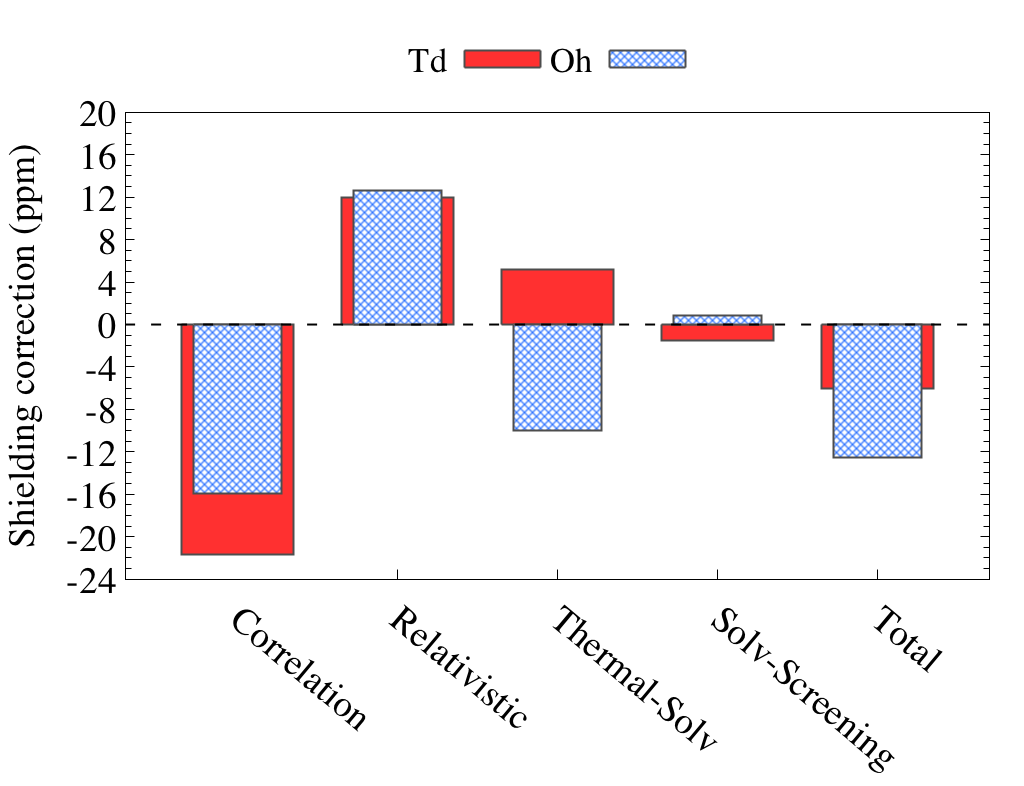}\par
\end{minipage}
\caption{Graphical representation of hierarchical shielding corrections (in ppm) to the HF/pCVQZ.TZ shielding constant of \al in \altetrahedral (red) and \aloctahedral (blue pattern). Electron correlation (at the CCSD/pCVQZ.TZ)  and relativistic corrections of \aloctahedral were extracted from Antu\v{s}ek et al.\cite{antuvsek2015absolute}}
\label{fig:corrections}
\end{figure}
Table \ref{table:nmr_correc} summarizes the shielding corrections of $\sigma_{Td}$ and Figure \ref{fig:corrections} shows a graphic comparison of the magnitude of the corrections for \altetrahedral and \aloctahedral species. Comparing the sign of the perturbation to the shielding constants (i.e deshielding (-) and shielding(+)) of both aluminum species it can be seen that, correlation and relativistic effects affected the shielding in a similar qualitative manner (both negative and positive deviations for correlation and relativistic effects respectively). Quantitatively, the electron correlation impact on the shielding of \altetrahedral is larger than the one on the \aloctahedral species (in absolute terms), whereas the relativistic effects are of very similar magnitude and would cancel out in a chemical shift calculation.  In contrast to the same-sign-effect behavior of the two previous effects, solvation corrections are of opposite sign. The total error due to averaged second-solvation effects and \altetrahedral/\aloctahedral molecular vibrations can be roughly approximated to the sum of the $\delta_{ss}$ small solvent-screening corrections and $\delta_{therm}$. Total solvation-induced error estimates become $+3.7$ and $-9$ ppm for \altetrahedral and \aloctahedral respectively. This additive approximation results in a total solvent-dynamical correction in \aloctahedral that agrees well with the one calculated, with a slightly different approach, by Antu\v{s}ek et al.\cite{antuvsek2015absolute}. Since solvation corrections are of opposite sign for the two \al species, they are summed in absolute terms, instead of cancelling out, toward the correction of the chemical shift of aluminate. Of all the effects on the nuclear magnetic shielding of \alisot studied here, \textit{combined solvent-dynamic effects become the largest correction to the aluminate chemical shift computation}.  

\section{Conclusions}
\label{sec:conclusions}
In this research, we present a high-level coupled cluster calculation of the \alisot NMR shielding in \altetrahedral. The estimated $^{27}$\altetrahedral isotropic shielding was 515.13 $\pm$ 5.3. Different contributions to the calculated shielding were evaluated in order to assess the pertinence of error cancellation in \altetrahedral chemical shift computations when \aloctahedral is used as the NMR reference. An initial assessment of the effects of level of theory (CCSD or B3LYP) and basis set choice during the geometry optimization steps revealed an influence of 2 to 5 ppm in the posterior HF/pCVQZ.TZ calculated shieldings (for both \al species). Performance and convergence behavior of a wide range of basis sets with respect to \alisot shielding in \aloctahedral and \altetrahedral was evaluated. Correlation consistent core-valence and pc-S Jensen basis showed monotonically decreasing convergence behavior of $\sigma(Al)$ towards the HF CBS limit as a function of increasing basis sets size. In this study, we fixed a discontinuity in the regular exponential trend of the Dunning basis set, which occurred at the cc-pVQZ level, by uncontracting one tighter p function and thus, we recommend this modification for any further computational study of Al NMR properties when this basis set type is chosen.
Hierarchical corrections to the HF/pCVQZ.TZ \al shielding in \altetrahedral and \aloctahedral were revealed. CCSD electron-correlation contribution has a deshielding effect, and constitutes 4.1\% of the Hartree-Fock value for the \altetrahedral, which is almost double the effect on \aloctahedral (2.5\%). Relativistic effects are significant ($\sim +12$ ppm) but of similar magnitude for both \al species, thus, this contribution cancels out in chemical shift computations. Solvation corrections to $\sigma_{NR}$, calculated as the sum of $\delta_{ss}$ and $\delta_{therm}$, were qualitatively and quantitatively system-specific. Total solvation-induced  error  estimates  were  $+3.7$  and $-9$ ppm for \altetrahedral and \aloctahedral respectively. Therefore, solvent-dynamic effects are the largest correction in \alisot chemical shift computations of aluminate. Including solvation effects, we obtained a value of 84.87 $\pm$ 6.7 for the coupled cluster $^{27}$\altetrahedral chemical shift. The computed \altetrahedral shielding constant can be used as the literature standard in the study of NMR-sensitive chemical phenomena around aluminate when cancellation of errors cannot be relied upon.

\begin{acknowledgments}
This work was supported by IDREAM (Interfacial Dynamics
in Radioactive Environments and Materials), an Energy
Frontier Research Center funded by the U.S. Department of
Energy (DOE), Office of Science, Basic Energy Sciences
(BES). E.M.B. is grateful for support from the Pacific Northwest National Laboratory (PNNL)- Washington State University (WSU) Distinguished Graduate Research Program (DGRP) Fellowship. Computations were performed at the Molecular Science Computing Facility at the Environmental Molecular Sciences Laboratory (EMSL) under proposal number 49771. EMSL (grid.436923.9) is a national scientific user facility sponsored by the DOE’s Office of Biological and Environmental Research located at PNNL. PNNL is a multiprogram national laboratory operated for DOE by Battelle Memorial Institute under Contract No. DE-AC06-76RLO-1830.
\end{acknowledgments}

\bibliography{main.bib}

\pagebreak
\widetext
\begin{center}
\textbf{\large Supplemental Materials:} \alisot NMR chemical shift of \altetrahedral from first principles. Assessment of error cancellation in NMR chemical shift computations in chemically distinct reference and targeted systems.
\end{center}
\setcounter{equation}{0}
\setcounter{figure}{0}
\setcounter{table}{0}
\setcounter{page}{1}
\makeatletter
\renewcommand{\thetable}{S\arabic{table}}
\renewcommand{\theequation}{S\arabic{equation}}
\renewcommand{\thefigure}{S\arabic{figure}}
\renewcommand{\thepage}{S\arabic{page}}

\begin{turnpage}
\begin{figure*}[htbp]
\centering
\begin{tabular}{@{}ccc@{}}
    \includegraphics[width=0.33\linewidth]{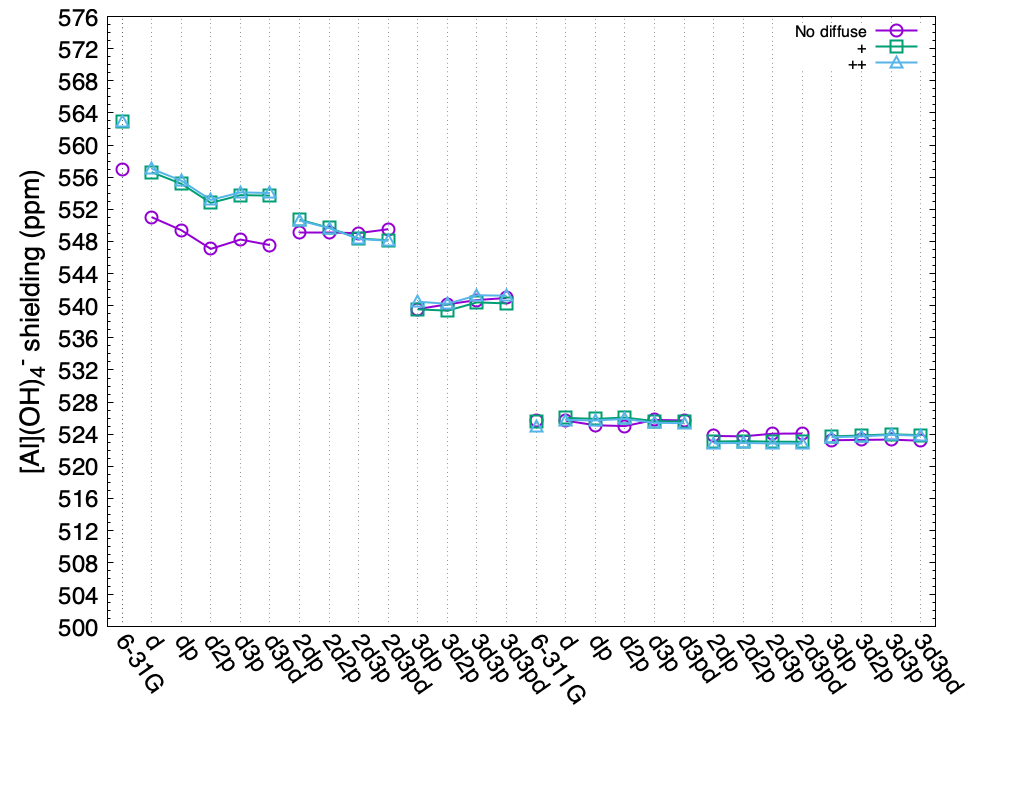} 
    \includegraphics[width=0.33\linewidth]{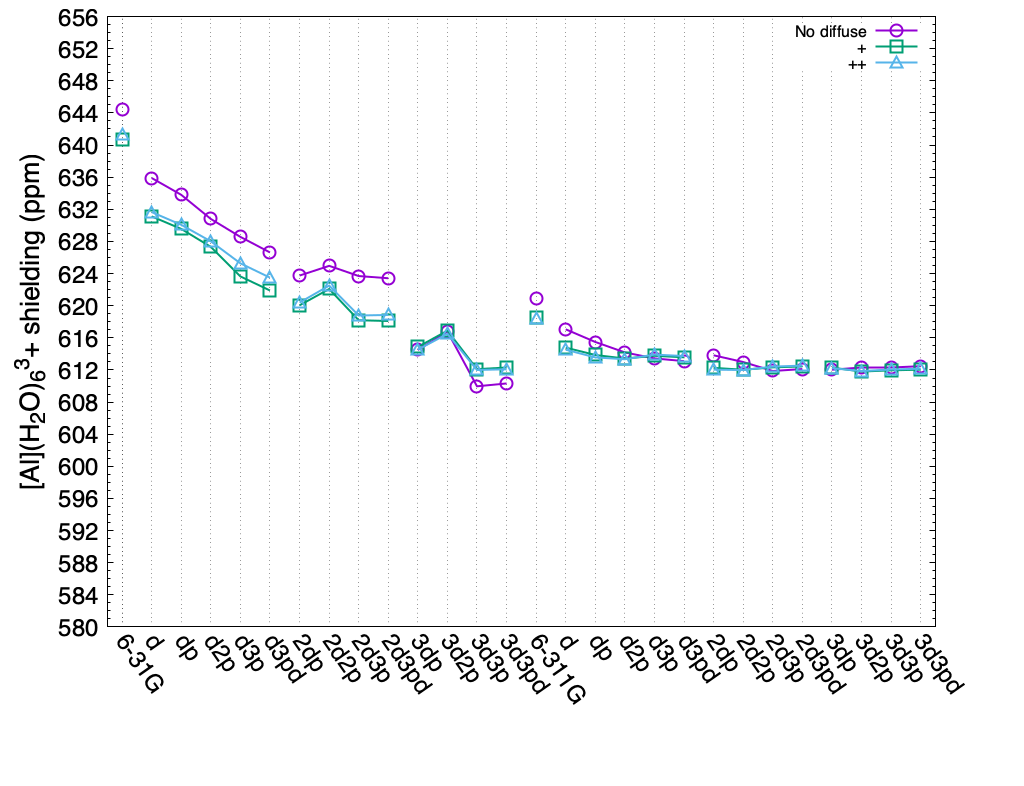} 
    \includegraphics[width=0.33\linewidth]{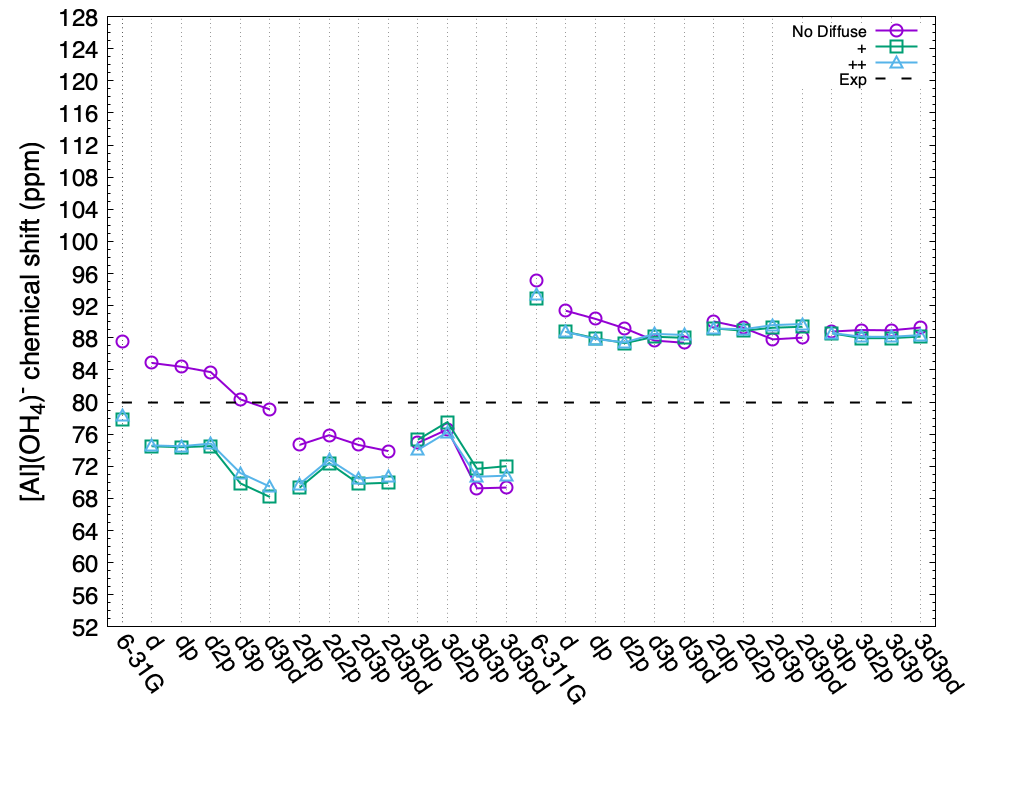} \\
    \includegraphics[width=0.33\linewidth]{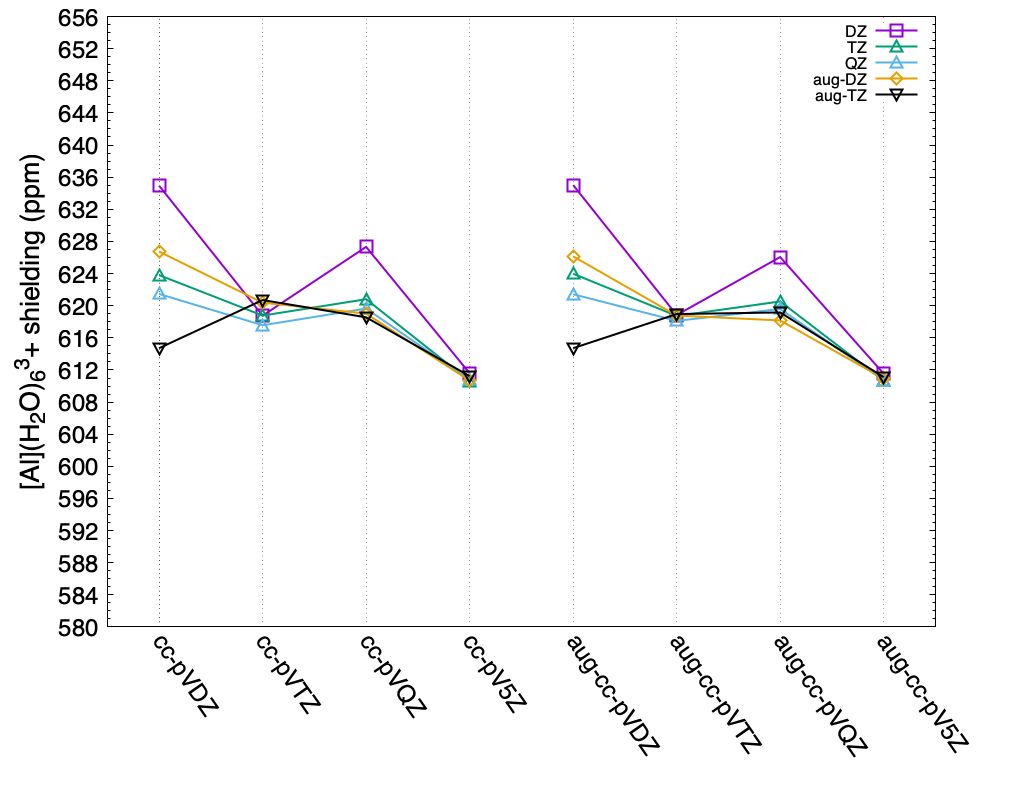} 
    \includegraphics[width=0.33\linewidth]{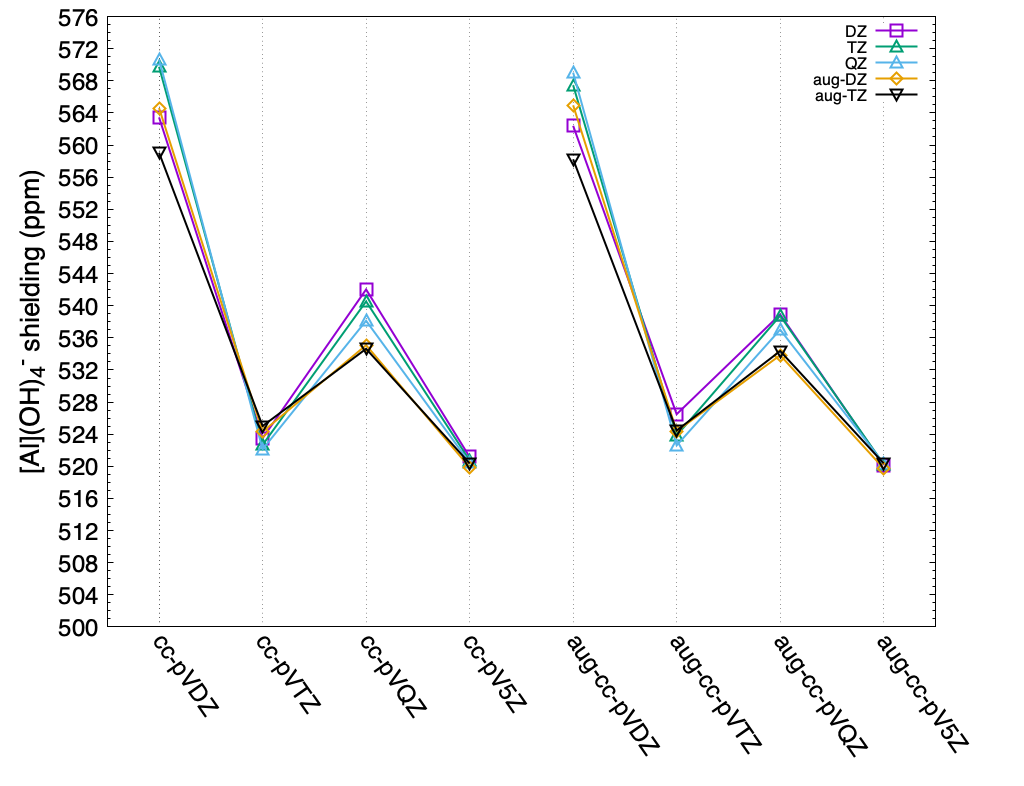} 
    \includegraphics[width=0.33\linewidth]{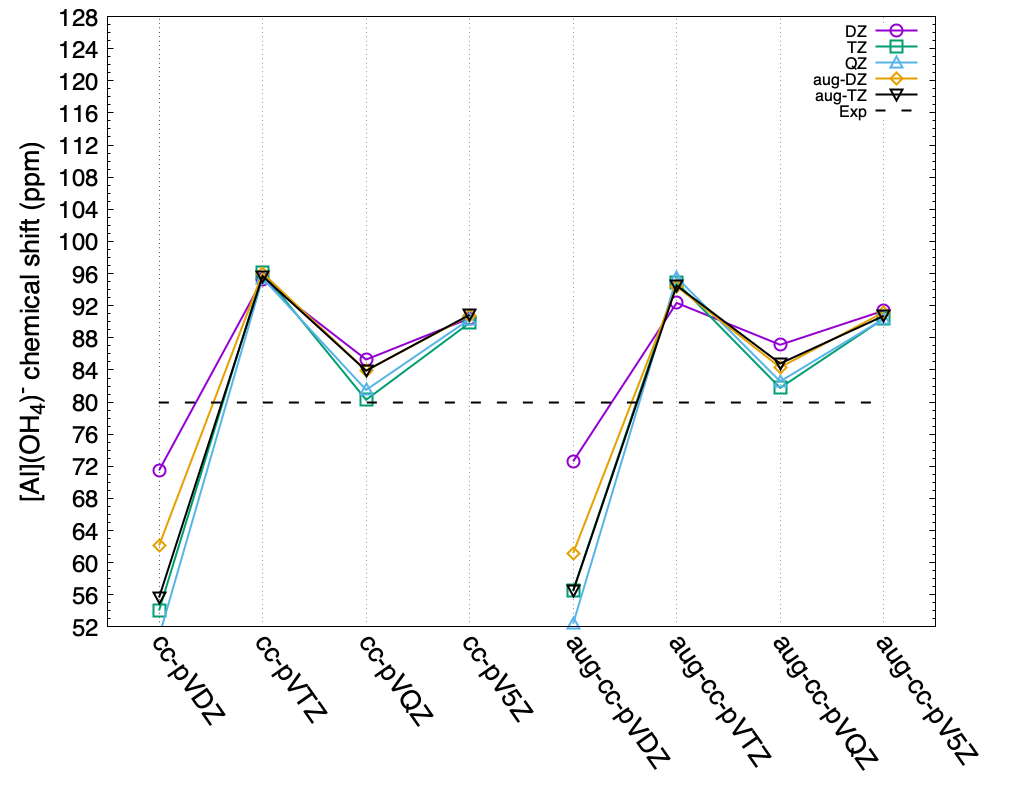} \\
\end{tabular}
\caption{HF GIAO Al NMR shielding convergence behavior as a function of basis sets type and quality. Gas phase \aloctahedral and \altetrahedral models optimized at CCSD/cc-pVTZ level}
\end{figure*}

\begin{figure*}[htbp]\ContinuedFloat
\centering
\begin{tabular}{@{}ccc@{}}
    \includegraphics[width=0.33\linewidth]{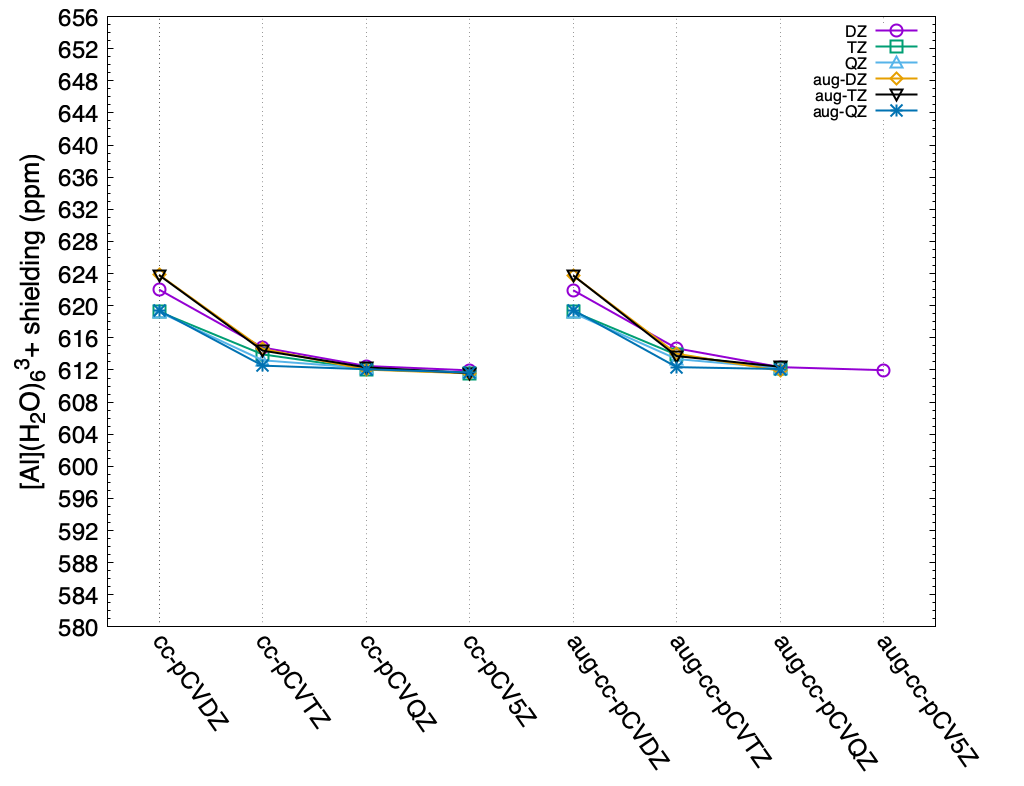} 
    \includegraphics[width=0.33\linewidth]{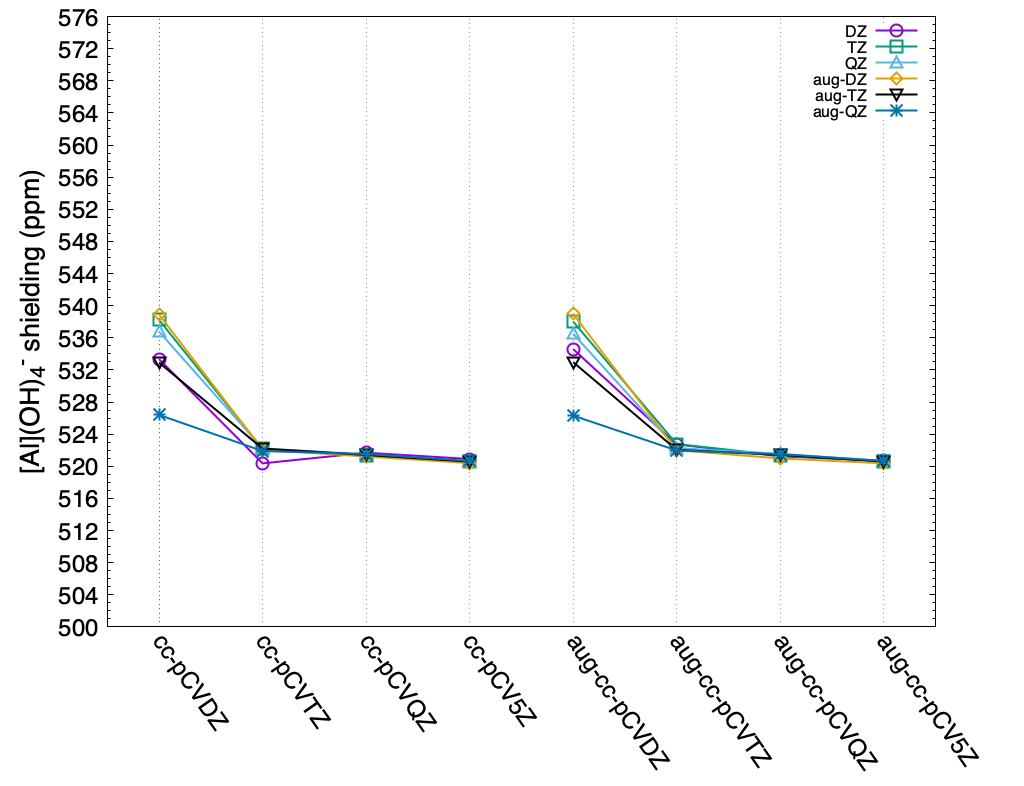} 
    \includegraphics[width=0.33\linewidth]{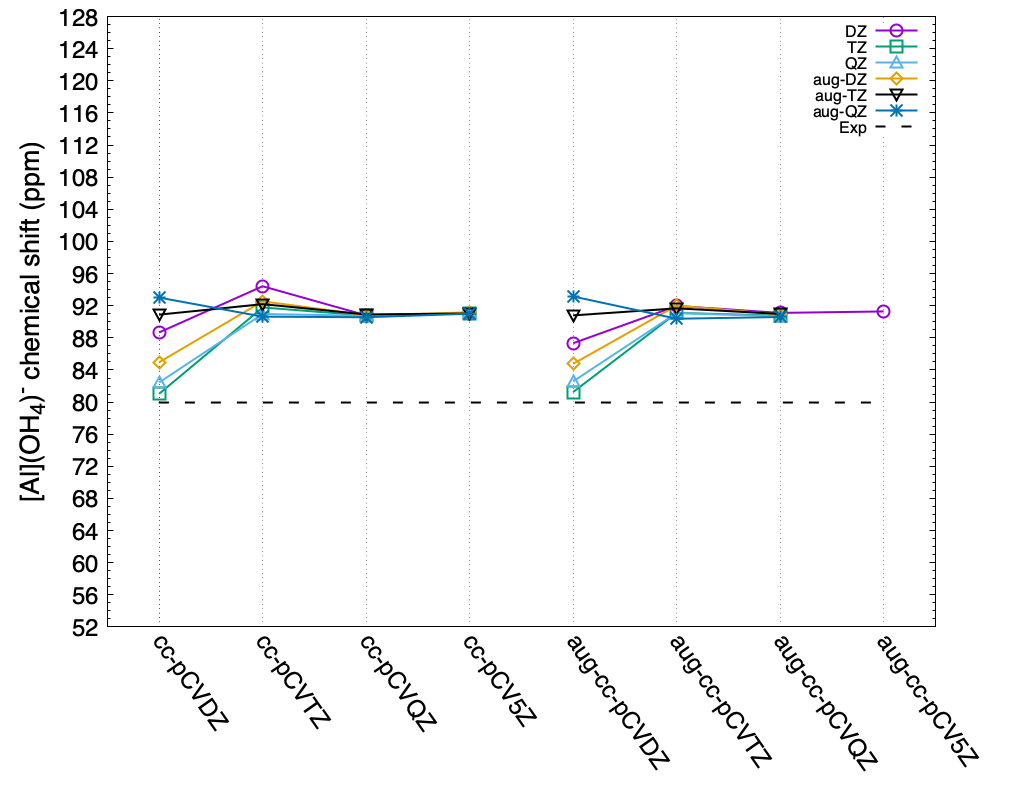} \\ 
    \includegraphics[width=0.33\linewidth]{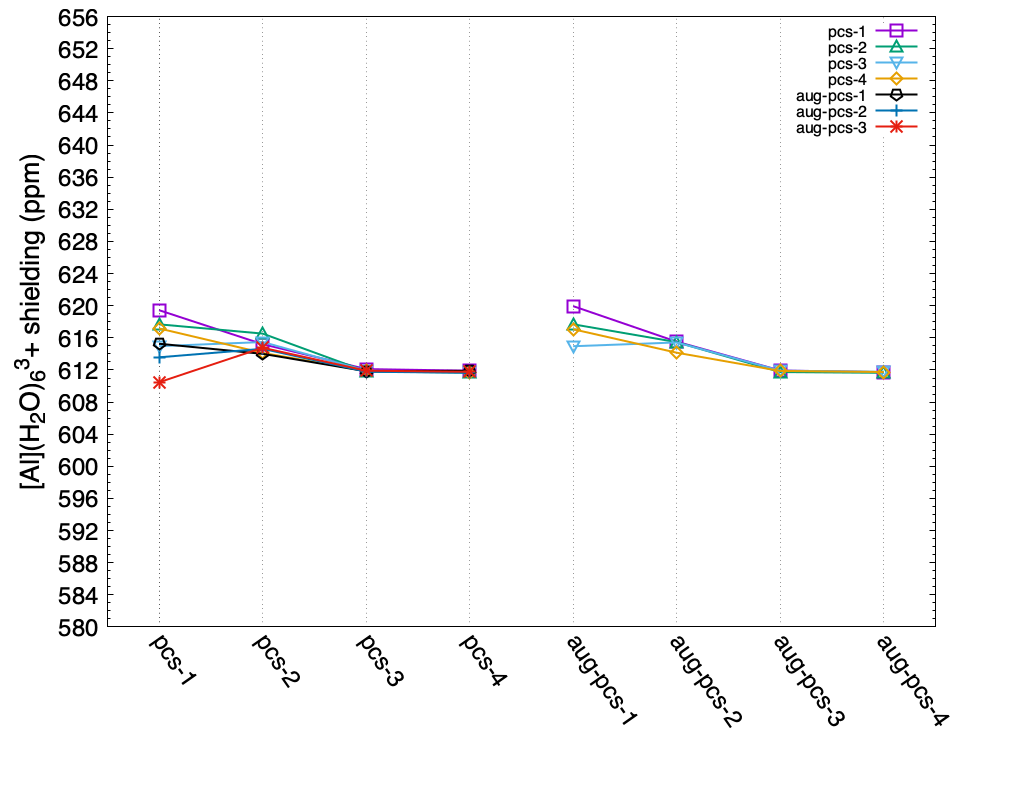} 
    \includegraphics[width=0.33\linewidth]{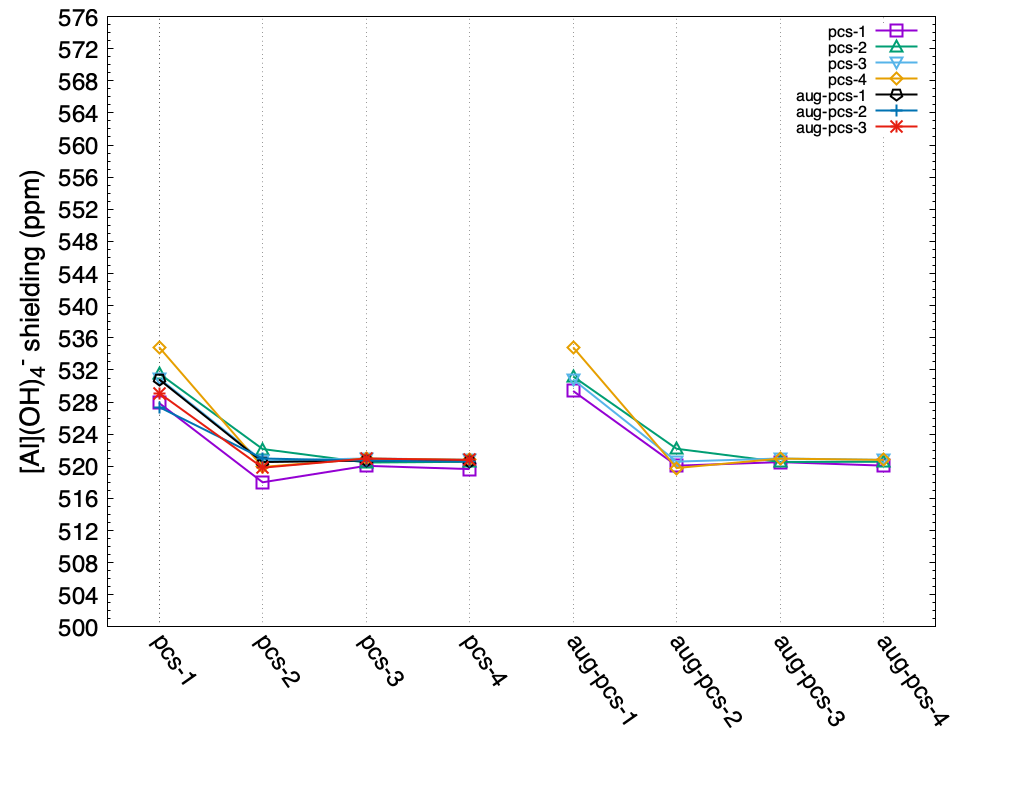} 
    \includegraphics[width=0.33\linewidth]{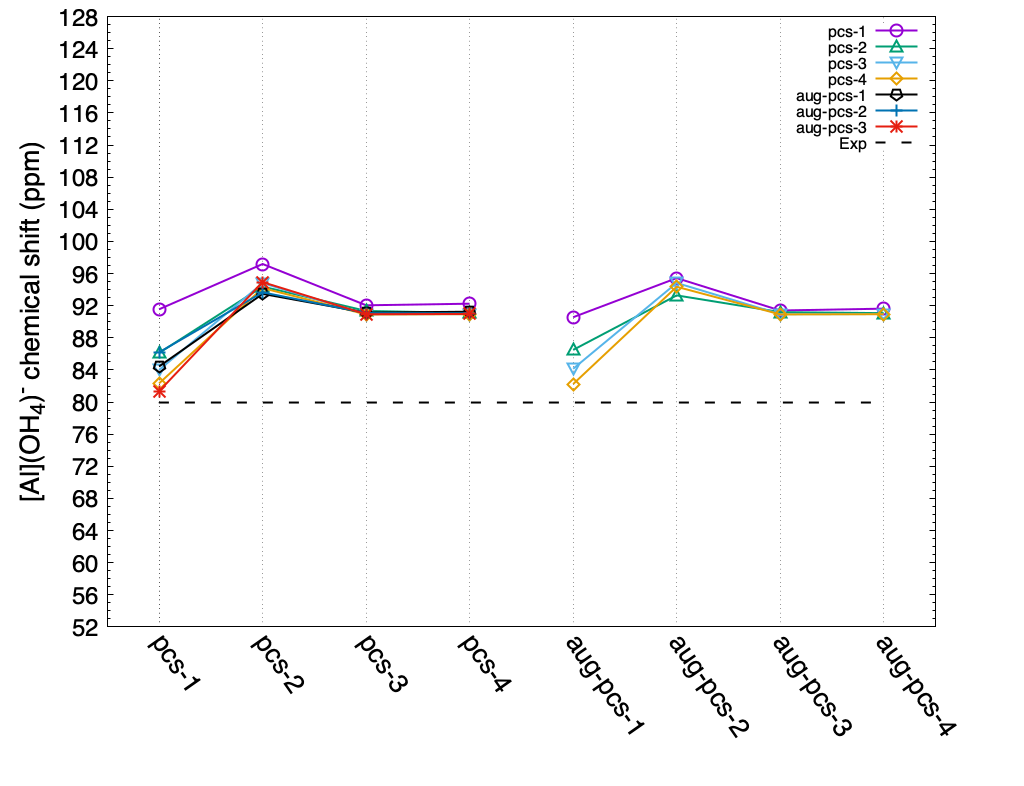} \\
\end{tabular}
\caption{HF GIAO Al NMR shielding convergence behavior as a function of standard and core-valence Dunning basis sets quality. Gas phase \aloctahedral and \altetrahedral models optimized at CCSD/cc-pVTZ level}
\label{fig:bs}
\end{figure*}
\end{turnpage}
\newpage

\begin{table}[!h]
\centering
\caption{Uncertainty associated with $\sigma_{NR}$ calculation. }\label{table:uncert}
\begin{tabular}{@{\extracolsep{4pt}}c c c c c}
\hline
\hline
Ion             &  Geometry uncertainty & Basis set incompleteness & Model Truncation & Total  \\ 
\cline{1-1} \cline{2-2} \cline{3-3} \cline{4-4} \cline{5-5}   
\altetrahedral  & -1.5 & -3.2 & 3.9 & 5.3 \\ 
\aloctahedral\footnote{$^{27}Al$ shielding values for \aloctahedral are taken from \textcite{antuvsek2015absolute}}   &1.3 & -2.3 & -3.1 & 4.1 \\
\hline
\hline
\end{tabular} 
\end{table}
\begin{table*}[h!]
\caption{DFT $^{27}\text{Al}$ NMR shielding constants (in ppm) calculated with the pCVQZ.TZ basis set combination. Geometries were optimized at the CCSD/cc-pVTZ level.}
\label{table:nmr_dft}
\begin{tabular}{l@{\hspace{1cm}} c@{\hspace{0.3cm}}c@{\hspace{0.5cm}} c@{\hspace{0.3cm}}c@{\hspace{0.5cm}} c@{\hspace{0.3cm}}c}
\hline
      & \multicolumn{2}{c}{$\sigma($\aloctahedral$)$} & \multicolumn{2}{c}{$\sigma($\altetrahedral$)$} &\multicolumn{2}{c}{$\delta($\altetrahedral$)$}  \\ 

\textbf{Functional} & DFT & $\Delta_{HF}$\footnote{$\Delta_{HF}$ is the difference between the value at the DFT level and the HF one for the property given in the column} & DFT & $\Delta_{HF}$ & DFT & $\Delta_{HF}$ \\
\cline{2-3} \cline{4-5} \cline{6-7} 
BLYP 	  	& 560.2721 &	-51.7395  & 461.3903  &	-59.7576  & 98.8818 &	8.0181 \\ 
B1LYP 		& 574.5055 &	-37.5061 	& 477.0484 	& -44.0995  &	97.4571 & 6.5934  \\
B3LYP 		& 571.5439 &	-40.4677 	& 473.6607 	& -47.4872  &	97.8832	&	7.0195 \\
BHandHLYP & 587.1734 & 	-24.8382  & 491.2658 	& -29.8821  &	95.9076	&	5.0439 \\
KT1 		  & 566.7078 &	-45.3038 	& 470.5253 	& -50.6226  &	96.1825	&	5.3188 \\
KT2 		  & 572.6338 &	-39.3778 	& 478.0450 	& -43.1029  &	94.5888	&	3.7251 \\
KT3 		  & 581.9601 &	-30.0515 	& 490.7505 	& -30.3974  &	91.2096	&	0.3459 \\
PBE 		  & 561.1035 &	-50.9081 	& 462.7419 	& -58.406  	& 98.3616	&	7.4979 \\
RPBE\footnote{Differences in shieldings between PBE and RPBE calculations are on the $10^{-7}$ ppm order. Hence their values in this Table are identical}	 	  & 561.1035 & 	-50.9081  & 462.7419  & -58.406   & 98.3616 &	7.4979 \\
revPBE 		& 566.4371 & 	-45.5745  & 469.9035  & -51.2444  & 96.5336 &	5.6699 \\
PBE0 		  & 575.7621 & 	-36.2495  &	478.9304  & -42.2175  & 96.8317 &	5.968 \\
\hline
\end{tabular} 
\end{table*}

\newpage

\begin{table*}[h!]
\caption{Gas phase models optimized at the CCSD/cc-pVTZ}
\centering
\label{table:gas_ccsd_xyz}
\begin{tabular}{c@{\hspace{1cm}} c@{\hspace{1cm}}c@{\hspace{1cm}}c@{\hspace{0.5cm}}}
\hline  
\altetrahedral & X & Y & Z \\
\cline{1-4} 
Al  &   -0.00001800  &  0.00001700  & -0.00000100 \\ 
O   &   -1.13253200  & -1.00763300  & -0.92949600 \\  
H   &   -0.76656400  & -1.87751100  & -1.07388300 \\  
O   &   -1.00776300  &  1.13267700  &  0.92920300 \\  
H   &   -1.87742100  &  0.76645500  &  1.07426700 \\  
O   &    1.00750700  & -1.13262000  &  0.92949100 \\  
H   &    1.87747700  & -0.76681400  &  1.07373400 \\  
O   &    1.13279300  &  1.00761600  & -0.92917500 \\  
H   &    0.76670700  &  1.87732200  & -1.07428700 \\ 
\hline
\end{tabular}
\bigskip

\begin{tabular}{c@{\hspace{1cm}} c@{\hspace{1cm}}c@{\hspace{1cm}}c@{\hspace{0.5cm}}}
\hline  
\aloctahedral & X & Y & Z \\
\cline{1-4} 
Al    &     -0.000020  &  0.000003  &  0.000010 \\ 
O     &     -0.572920  &  1.702532  &  0.688165 \\ 
O     &      1.376428  & -0.079552  &  1.341460 \\ 
O     &     -1.216391  & -0.891938  &  1.193856 \\ 
O     &      0.572991  & -1.702490  & -0.688190 \\ 
O     &     -1.376436  &  0.079555  & -1.341449 \\ 
O     &      1.216359  &  0.891903  & -1.193879 \\ 
H     &     -0.189431  &  2.183634  &  1.438128 \\ 
H     &     -1.302318  &  2.246526  &  0.351990 \\ 
H     &      1.299863  & -0.463662  &  2.228906 \\ 
H     &      2.282130  &  0.258508  &  1.261545 \\ 
H     &     -1.813537  & -0.472269  &  1.832812 \\ 
H     &     -1.349287  & -1.849329  &  1.275731 \\ 
H     &      0.189003  & -2.184022  & -1.437621 \\ 
H     &      1.302740  & -2.246197  & -0.352314 \\ 
H     &     -1.299981  &  0.464150  & -2.228696 \\ 
H     &     -2.282003  & -0.258915  & -1.261751 \\ 
H     &      1.812682  &  0.472216  & -1.833588 \\  
H     &      1.350149  &  1.849237  & -1.274970 \\ 
\hline
\end{tabular}
\end{table*}

\newpage

\begin{table*}[h!]
\caption{Gas phase models optimized at the B3LYP/cc-pVTZ}
\centering
\label{table:gas_b3lyp_xyz}
\begin{tabular}{c@{\hspace{1cm}} c@{\hspace{1cm}}c@{\hspace{1cm}}c@{\hspace{0.5cm}}}
\hline  
\altetrahedral & X & Y & Z \\
\cline{1-4} 
Al  &  -0.000019 &    0.000015 &    -0.000001 \\
O   & -1.059364  &   -1.086377 &    -0.933790 \\
H   & -0.644690  &   -1.937918 &    -1.079037 \\
O   & -1.086493  &   1.059502  &   0.933516 \\
H   & -1.937839  &   0.644647  &   1.079377 \\
O   & 1.086252   &  -1.059452  &   0.933797 \\ 
H   & 1.937883   &  -0.644934  &   1.078951 \\ 
O   & 1.059608   &  1.086359   &  -0.933506 \\ 
H   & 0.644866   &  1.937751   &  -1.079418 \\
\hline
\end{tabular}
\bigskip

\begin{tabular}{c@{\hspace{1cm}} c@{\hspace{1cm}}c@{\hspace{1cm}}c@{\hspace{0.5cm}}}
\hline  
\aloctahedral & X & Y & Z \\
\cline{1-4} 
Al &   -0.000002 &   0.000000 &    0.000002 \\
O  &  1.412618   & -0.718838  &   -1.110367 \\
O  &  0.359361   & -1.355213  &   1.332487 \\
O  &  1.274621   & 1.178386   &  0.855114 \\
O  &  -1.412617  &  0.718840  &   1.110366 \\
O  &  -0.359366  &  1.355213  &   -1.332485 \\
O  &  -1.274618  &  -1.178386 &    -0.855116 \\
H  &  1.979462   & -1.483152  &   -0.901743 \\
H  &  1.689986   & -0.386189  &   -1.982962 \\
H  &  0.980028   & -1.282331  &   2.079841 \\
H  &  -0.050744  &  -2.237289 &    1.385847 \\
H  &  2.227692   & 1.237893   &  0.662152 \\ 
H  &  1.084319   & 1.824018   &  1.559426 \\
H  &  -1.979441  &  1.483171  &   0.901749 \\ 
H  &  -1.690003  &  0.386178  &   1.982950 \\
H  &  -0.979974  &  1.282300  &   -2.079885 \\
H  &  0.050698   & 2.237310   &  -1.385811 \\
H  &  -2.227708  &  -1.237833 &    -0.662225 \\
H  &  -1.084287  &  -1.824093 &    -1.559351 \\
\hline
\end{tabular}
\end{table*}

\newpage

\begin{longtable}{c@{\hspace{1cm}} c@{\hspace{1cm}}c@{\hspace{1cm}}c@{\hspace{0.5cm}}}
\hline  
\superaltetrahedral & X & Y & Z \\
\cline{1-4}
Al      &           0.06235600  & -0.05206200 &  -1.54115100\\
O       &           0.00011500  & -0.22892400 &   0.27181000\\
O       &           1.58266800  &  0.70256500 &  -2.00913500\\
O       &          -0.09262400  & -1.71567900 &  -2.11481600\\
O       &          -1.29420000  &  0.95224000 &  -2.06390500\\
O       &           0.15757100  &  1.73730100 &   2.56859000\\
O       &          -1.97343900  & -1.65344100 &   1.42845200\\
O       &           2.49833100  &  2.42960500 &   1.63862100\\
O       &          -2.20358400  &  2.61952700 &   1.62267900\\
O       &           1.67295200  & -1.97250300 &   1.44803500\\
O       &           3.79279100  &  0.03541100 &   1.29293200\\
O       &          -3.93888800  &  0.30670500 &   1.25864100\\
O       &          -0.38487200  & -3.15604200 &   3.02508600\\
O       &           3.87881500  & -0.62660400 &  -1.39264800\\
O       &           1.53971200  &  3.44452600 &  -0.82680300\\
O       &          -2.48345800  & -2.63873700 &  -1.23680300\\
O       &           2.35965100  & -3.10688900 &  -0.93266100\\
O       &          -1.11613600  &  3.39323600 &  -0.99075400\\
O       &          -3.88809800  & -0.39561700 &  -1.49388200\\
H       &           0.08373000  &  0.57116800 &   0.80485300\\
H       &           1.69319700  &  1.62549500 &  -1.74854300\\
H       &           0.64731900  & -2.30441200 &  -1.92604800\\
H       &          -2.18696100  &  0.59724800 &  -1.96879500\\
H       &           1.07511300  &  2.02552500 &   2.26553400\\
H       &           0.27414900  &  0.93922600 &   3.09036900\\
H       &          -1.27218900  & -1.10441200 &   0.98956500\\
H       &          -2.72979900  & -1.03580000 &   1.53147600\\
H       &           3.00527400  &  1.59754600 &   1.47258000\\
H       &           2.28769300  &  2.80768500 &   0.76543100\\
H       &           1.07419100  & -1.31382200 &   1.00485500\\
H       &           1.09402800  & -2.43860900 &   2.08127000\\
H       &          -3.40099000  &  1.11253700 &   1.34027100\\
H       &          -4.04743400  &  0.16301900 &   0.30064700\\
H       &          -1.35363700  &  2.29486400 &   2.00642600\\
H       &          -2.52029100  &  3.29432100 &   2.22843900\\
H       &           3.17472000  & -0.65769400 &   1.57391300\\
H       &           3.91935900  & -0.15534400 &   0.33385000\\
H       &          -1.06095600  & -2.59885700 &   2.57064200\\
H       &          -0.53971800  & -4.03831000 &   2.67666300\\
H       &           3.08898400  & -0.11348300 &  -1.69507900\\
H       &           3.55608400  & -1.53985200 &  -1.36694500\\
H       &          -1.60230100  & -2.40002800 &  -1.62797500\\
H       &          -2.34556900  & -2.52736000 &  -0.27908200\\
H       &          -1.24357300  &  2.52750900 &  -1.46835600\\
H       &          -1.53436300  &  3.24658900 &  -0.12766200\\
H       &           0.54256900  &  3.48852700 &  -0.86539200\\
H       &           1.86142500  &  4.18223600 &  -1.34951000\\
H       &           2.63813200  & -4.02271200 &  -0.86460500\\
H       &           2.13212900  & -2.81805100 &  -0.01631500\\
H       &          -3.42019500  & -1.27906100 &  -1.46846700\\
H       &          -4.55921500  & -0.46231300 &  -2.17655400\\
\hline
\end{longtable}

\newpage
\begin{longtable}{c@{\hspace{1cm}} c@{\hspace{1cm}}c@{\hspace{1cm}}c@{\hspace{0.5cm}}}
\hline  
\suprealoctahedral  & X & Y & Z \\
\cline{1-4} 
Al &   0.000047 &   -0.000204 &   -0.000050 \\
O  &  0.583203  &  -1.467539  &  -1.085315 \\
O  &  0.978540  &  1.238539   & -1.086284 \\
O  &  -1.562262 &   0.227629  &  -1.085915 \\
O  &  -0.583048 &   1.467195  &  1.085179 \\
O  &  -0.978632 &   -1.238805 &   1.086134 \\
O  &  1.562342  &  -0.227981  &  1.085768 \\
O  &  3.218447  &  -1.903750  &  -1.355471 \\
O  &  -1.019809 &   -3.624369 &   -1.325135 \\
O  &  0.037850  &  3.738691   & -1.356546 \\
O  &  3.648133  &  0.930499   & -1.325796 \\
O  &  -3.259111 &   -1.836321 &   -1.354500 \\
O  &  -2.631918 &   2.692894  &  -1.324377 \\
O  &  -3.218253 &   1.904388  &  1.355205 \\
O  &  1.019618  &  3.624176   & 1.324532 \\
O  &  -0.037697 &   -3.738864 &   1.355917 \\
O  &  -3.648161 &   -0.930145 &   1.327178 \\
O  &  3.258448  &  1.836610   & 1.355398 \\
O  &  2.632394  &  -2.692749  &  1.323975 \\
H  &  1.530058  &  -1.647989  &  -1.316651 \\
H  &  0.027297  &  -2.254788  &  -1.308463 \\
H  &  0.660955  &  2.148680   & -1.317390 \\
H  &  1.937806  &  1.150533   & -1.310899 \\
H  &  -2.192544 &   -0.501931 &   -1.315962 \\
H  &  -1.966476 &   1.102552  &  -1.308577 \\
H  &  -1.529886 &   1.647515  &  1.316627 \\
H  &  -0.027207 &   2.254526  &  1.308242 \\
H  &  -0.661179 &   -2.149002 &   1.317237 \\
H  &  -1.937954 &   -1.150774 &   1.310552 \\
H  &  2.192463  &  0.501727   & 1.315798 \\
H  &  1.966847  &  -1.102826  &  1.308325 \\
H  &  3.633899  &  -1.021406  &  -1.398448 \\
H  &  3.633726  &  -2.439014  &  -2.044476 \\
H  &  -0.933985 &   3.657067  &  -1.399117 \\
H  &  0.293218  &  4.365188   & -2.046458 \\
H  &  -2.702831 &   -2.637280 &   -1.397992 \\
H  &  -3.930012 &   -1.927571 &   -2.043877 \\
H  &  -0.791088 &   -4.049875 &   -0.478657 \\
H  &  -0.960113 &   -4.306092 &   -2.007794 \\
H  &  3.900303  &  1.341661   & -0.478813 \\
H  &  4.209299  &  1.323838   & -2.007482 \\
H  &  -3.113792 &   2.706478  &  -0.477232 \\
H  &  -3.253423 &   2.981333  &  -2.006178 \\
H  &  -3.634167 &   1.022326  &  1.399628 \\
H  &  -3.633144 &   2.440871  &  2.043502 \\
H  &  0.934102  &  -3.657101  &  1.398309 \\
H  &  -0.292785 &   -4.365488 &   2.045834 \\
H  &  2.701792  &  2.637296   & 1.398896 \\
H  &  3.928798  &  1.927883   & 2.045316 \\
H  &  0.791340  &  4.048917   & 0.477547 \\
H  &  0.959324  &  4.306519   & 2.006520 \\
H  &  -3.901478 &   -1.341248 &   0.480505 \\
H  &  -4.208713 &   -1.323248 &   2.009505 \\
H  &  3.114760  &  -2.706067  &  0.477137 \\
H  &  3.253524  &  -2.981334  &  2.006060 \\
\hline
\end{longtable}
\newpage

\end{document}